\documentclass[letterpaper,twocolumn,10pt]{article}

\usepackage{titlesec}

\usepackage[hyphens]{url}
\usepackage{hyperref}
\usepackage{xifthen}
\usepackage[usenames,svgnames,table]{xcolor}

\usepackage{usenix2019_v3}

\usepackage[utf8]{inputenc}%
\usepackage{xspace}
\usepackage{color}
\usepackage{latexsym}
\usepackage{hyperref}
\usepackage{comment}
\usepackage{enumerate}
\usepackage{mdwlist}

\usepackage[pdftex]{graphicx}
\usepackage{grffile} %
\graphicspath{{figures/}{../src/shadow_perf_analysis/}{../src/shadow_measurement_analysis/}}
\DeclareGraphicsExtensions{.pdf}

\usepackage[numbers,square,sort&compress]{natbib}
\usepackage{amssymb, amsthm, amsfonts, latexsym, amsmath}
\usepackage{wasysym} %
\usepackage[font=small,labelfont=bf]{caption}
\usepackage{soul}
\usepackage{rotating}
\usepackage{multirow}
\usepackage{multicol}

\usepackage[size=footnotesize,subrefformat=parens]{subcaption}
\usepackage{float}

\usepackage{booktabs}

\usepackage{threeparttable}

\usepackage{versions}
\includeversion{full}
\excludeversion{conference}

\newcounter{mn}
\newcommand{\superscript}[1]{\ensuremath{{}^{\textrm{\scriptsize #1}}}}

\newcommand{\mntext}[1]{\colorbox{SkyBlue}{\begin{color}{black}#1\end{color}}}
\newcommand{\mn}[2][]{{\tiny\superscript{\mntext{\arabic{mn}}}}\marginpar{\scriptsize{
  \ifthenelse{\isempty{#1}}
  {\mntext{\parbox{0.98\marginparwidth}{\superscript{\arabic{mn}}~\raggedright{#2}}}}
  {\mntext{\parbox{0.98\marginparwidth}{\superscript{\arabic{mn}}#1:~\raggedright{#2}}}}
}}\stepcounter{mn}}

\newcommand{\tool}{FlashFlow\xspace}
\newcommand{\mytitle}{\tool: A Secure Speed Test for Tor}

\hypersetup{
     bookmarks=true,
     colorlinks=true,
     linkcolor=black,
     citecolor=black,
     filecolor=black,
     urlcolor=black,
     pdfauthor = {Matthew Traudt, Rob Jansen, and Aaron Johnson}, %
     pdftitle = {\mytitle},
     pdfkeywords = {Tor, security, anonymity, performance, measurement, bandwidth}
}

\PassOptionsToPackage{hyphens}{url}
\makeatletter
\g@addto@macro{\UrlBreaks}{\UrlOrds}
\makeatother

\newcommand\pagebudget[1]{~\textrm{}}

\definecolor{Red}{rgb}{1,0,0}
\definecolor{Orange}{rgb}{1,0.5,0}
\definecolor{Green}{rgb}{0,1,0}
\definecolor{Blue}{rgb}{0,0,1}
\definecolor{Magenta}{rgb}{1,0,1}
\definecolor{DarkGreen}{rgb}{0.0, 0.5, 0.0}

\renewcommand{\paragraph}[1]{\noindent\textbf{#1:}}

\newcommand{\Section}[1]{\section{#1}}
\newcommand{\Subsection}[1]{\subsection{#1}}

\titlespacing\section{0pt}{6pt plus 4pt minus 2pt}{6pt plus 2pt minus 2pt}
\titlespacing\subsection{0pt}{6pt plus 4pt minus 2pt}{6pt plus 2pt minus 2pt}
\titlespacing\subsubsection{0pt}{6pt plus 4pt minus 2pt}{6pt plus 2pt minus 2pt}

\begin{document}

\date{}

\author{
{\rm Matthew Traudt} \hspace{25mm} {\rm Rob Jansen} \hspace{25mm} {\rm Aaron Johnson}\\
{U.S. Naval Research Laboratory}\\
{\rm \{matthew.traudt, rob.g.jansen, aaron.m.johnson\}@nrl.navy.mil}
} %

\title{\mytitle}

\maketitle

\begin{abstract}
The Tor network uses a measurement system to estimate its relays' forwarding
capacity and to balance traffic among them. This system has been shown to be
vulnerable to adversarial manipulation. Moreover, its accuracy and effectiveness
in benign circumstances has never been fully quantified. We first obtain such a
quantification by analyzing Tor metrics data and performing experiments on the
live network. Our results show that Tor currently underestimates its true
capacity by about 50\% and improperly balances its traffic by 15–25\%. Then, to
solve the problems with security and accuracy, we present \tool, a system to
measure the capacity of Tor relays. Our analysis shows that \tool limits a
malicious relay to obtaining a capacity estimate at most 1.33 times its true
capacity. Through realistic Internet experiments, we find that \tool measures
relay capacity with $\ge$89\% accuracy 95\% of the time. Through simulation, we
find that \tool can measure the entire Tor network in less than 5 hours using 3
measurers with 1~Gbit/s of bandwidth each.
Finally, simulations using \tool{} for load balancing shows that, compared to TorFlow, network weight error decreases by 86\%,
while the median of 50\,KiB, 1\,MiB, and 5\,MiB transfer times decreases
by 15\%, 29\%, and 37\%, respectively. %
Moreover, \tool yields more consistent client performance:
the median rate of transfer timeouts decreases by 100\%, %
while the standard deviation of 50\,KiB, 1\,MiB, and 5\,MiB transfer times decreases
by 55\%, 61\%, and 41\%, respectively. %
We also find that the performance improvements increase relative to
TorFlow as the total client-traffic load increases, demonstrating that
\tool is better suited to supporting network growth.
\end{abstract}

\Section{Introduction \pagebudget{1.5}}

Tor~\cite{tor} is the most popular system on the Internet for anonymous
communication. Tor is currently comprised of about 6,500 geographically diverse
volunteer-operated proxy \textit{relays} transferring nearly 200 Gbit/s in
aggregate traffic from between 2~million~\cite{tormetrics} and
8~million~\cite{tor-usage-imc} daily active users. Tor has seen significant
growth recently, nearly doubling the amount of traffic it forwards in the last
two years~\cite{tormetrics}.

Tor uses a load-balancing system called TorFlow~\cite{perry2009torflow} to
balance load from its millions of users across its thousands of relays. The goal
of TorFlow is to equalize Tor performance across all clients, regardless of
which relays they use. It receives bandwidth self-measurements from relays and
also makes active measurements of download speeds through each relay. It then
computes per-relay weights by multiplying the self-measured bandwidths by their
actively measured speed relative to the average. Clients choose relays for their
circuits with probabilities proportional to these weights.

Previous work has shown that TorFlow is insecure. A malicious relay can increase
the fraction of traffic it can observe beyond the fraction of Tor bandwidth it
provides~\cite{thill-thesis,oakland2013-trawling,low-resource-attack,peerflow-pets2017},
increasing its ability to deanonymize Tor users using a traffic correlation
attack~\cite{ccs2013-usersrouted,deepcor-ccs2018}. A main reason for its
vulnerability is that it trusts relays to accurately self-report their observed
capacity. Also, TorFlow's active measurements are supposed to occur concurrently
with normal client traffic, but a malicious relay can detect its measurement and
throttle client traffic to increase its measured speed.

In addition to its insecurity, TorFlow has not been demonstrated to be accurate
even when not under attack. A relay estimates its capacity using the maximum
amount of throughput it is able to sustain for any 10 second period over each of
the last 5 days~\cite[\S~2.1.1]{tor-protocol-spec}. However, a relay that is
consistently under-utilized may never produce an accurate self-estimate of its
capacity, leading TorFlow to produce lower weights for that relay than it
should. Moreover, the active measurements depend on client traffic and the speed
of other relays randomly chosen for the same measurement circuits, potentially
leading to suboptimal and variable weights. Inaccurate weights reduce client
performance by improperly balancing load. Moreover, inaccurate capacity
estimates make it more difficult to understand how to spend research and
development effort on improving the network. For example, obtaining funding to
improve Tor scalability is more challenging without understanding the current
limits of the network. Improper network management also complicates relay
recruitment and retention, and may dissuade the development of incentive
schemes~\cite{lira-ndss13,ccs10-braids,incentives-fc10,Moore2011tortoise}.

\looseness-1
We explore the error and inconsistency in Tor's estimated relay capacities and
weights using Tor metrics data~\cite{tormetrics} and an active measurement
experiment. Our analysis of 11 years of data shows that 25\% of relays have a
mean capacity error of 49\% or greater, that total network capacity error has
reached as high as 60\%, and that relay capacity estimates vary by 82\% or
greater for 25\% of relays. The analysis also shows median load balancing errors
between 15\% and 25\% over time. Our measurement experiment on Tor further
indicates that relays significantly under-estimate their own capacity, and the
network capacity as a whole is underestimated by about 50\%.

We present \tool to solve these problems. \tool is a system designed to
securely, accurately, and quickly measure the capacity of relays in the Tor
network. In addition to providing weights for load balancing, the capacity
measurements allow Tor to accurately assess the network's resources and plan for
the future.

The need for security heavily influences the design choices of \tool. We cannot
make use of measurement approaches that are vulnerable to manipulation, such as
packet pairs~\cite{prasad2003bandwidth}. Previously proposed systems attempt to
measure Tor surreptitiously~\cite{perry2009torflow,andre2018smartor} or to
securely aggregate passive observations made by many
relays~\cite{eigenspeed,peerflow-pets2017}. \tool takes a new approach to this
problem by using separate measurement teams that attempt to actively utilize the
\emph{full} capacity of relays. This approach improves security as it requires
the direct demonstration of a relay's capacity rather than relying on an
indirect measurement that may be falsifiable. It also yields higher accuracy, as
the traffic is actively generated to determine the relay's limit, with the
normal client traffic carefully reduced to limit its impact on the result
without excessively reducing client performance. \tool additionally aggregates
results from multiple measurers in order to accurately measure the
highest-bandwidth relays in Tor.

We implement \tool and conduct extensive experiments in a lab setting, on the
Internet, and in simulation. With our suggested parameter settings, \tool limits
a malicious relay to obtaining a capacity estimate of at most 1.33 times its
true capacity. Through Internet experiments across a range of geographic
locations, we find that \tool is able to measure a target relay with a capacity
ranging from 10 Mbit/s to 1 Gbit/s to within 11\% of ground truth in 30 seconds
95\% of the time (or within 20\% of ground truth 99.8\% of the time). Through
simulation, we find that \tool can measure the entire Tor network in less than 5
hours using 3 measurers each with 1 Gbit/s of bandwidth.
Through private Tor network simulations in
Shadow, we find that \tool reduces network
weight error by 86\%. The resulting improvement in load balancing
reduces transfer times for \textit{all} tested transfer sizes: the median of
50\,KiB, 1\,MiB, and 5\,MiB transfer times decreases by 15\%, 29\%,
and 37\%, respectively. \tool also yields more \textit{consistent} client
performance: the median rate of transfer timeouts decreases by 100\%,
while the standard deviation of 50\,KiB, 1\,MiB, and 5\,MiB transfer
times decreases by 55\%, 61\%, and 41\%, respectively.
Finally, we find that the performance improvements increase further as the
total client-traffic load increases, demonstrating that \tool is
better suited to supporting Tor network growth than is TorFlow.
\Section{Background \pagebudget{0.5}}
\label{sec:background}

\paragraph{Overview}
As of August 2019, the Tor network includes about 6,500
\textit{relays} that forward a combined 200 Gbit/s of Tor traffic, and
9 \textit{Directory Authorities} (DirAuths) that act as trust anchors for the
distribution of network information to Tor users. When new relays join
the network, they publish their public key and network address
to the DirAuths, who then verify reachability and
validate Tor protocol support. A voting process occurs every hour,
after which the DirAuths add valid relays to a
\textit{network consensus} document signed by all authorities
and distributed to all Tor clients and relays. The consensus document
stores information about all available relays and is required for new
clients to use Tor. New relays that appear in a consensus are not used
until their performance has been measured by a majority of the 6
\textit{Bandwidth Authorities} (BWAuths) that participate in Tor's load
balancing system.

\paragraph{TorFlow} \label{sec:bwbackground}
Each Bandwidth Authority runs the TorFlow~\cite{perry2009torflow}
relay-measurement tool to measure the relative performance of relays in the Tor
network over time. TorFlow conducts performance measurements of Tor relays by
creating 2-hop Tor circuits through them and downloading one of a set of
13~fixed-sized files ($2^{i}$ KiB for $i \in \{4,\ldots,16\}$) from a known
destination through each circuit. Every hour, TorFlow
aggregates the latest relay measurements and produces a load-balancing
\emph{weight} for each relay. %

To assist in balancing load across relays, TorFlow attempts to produce larger
weights for relays that can better handle Tor traffic. To compute the
weights, TorFlow relies on two data sources.
First, TorFlow uses each relay's self-reported bandwidth information
that is published every 18~hours in a \textit{server descriptor}. This
information includes any rate limit set by the relay
(e.g., with the \textsf{BandwidthRate} and \textsf{BandwidthBurst}
options~\cite{tor-manual}), as well
as its \textit{observed bandwidth}, which is the highest Tor throughput that
the relay was able to sustain for any 10-second period during the last
5 days~\cite[\S2.1.1]{tor-protocol-spec}. From this information, TorFlow computes the relay's
\textit{advertised bandwidth} as the minimum of the observed bandwidth
and any rate limit set by the relay.
Second, TorFlow uses the results of its own measurements
to compute for each relay a ratio of the measurement speed of the
relay to the mean measurement speed of all relays in the network.
Finally, TorFlow computes a weight for each relay by multiplying the
computed speed ratio for that relay by its advertised bandwidth.

\paragraph{Load Balancing}
The TorFlow weights are collected and reported to the Directory
Authorities, added to the following network consensus, and distributed
to clients. Tor clients then use the normalized weights as probabilities when
selecting relays for theirs path through the Tor network in an attempt to
balance user load across relays. To use Tor, a client creates a \emph{circuit}
through a sequence of three relays, over which a TCP connection can then be made
to any Internet host. Communication \emph{cells} of a fixed 514-byte length are
sent through the circuit and are encrypted (or decrypted, depending on the
direction) by each relay using a key exchanged with the client during circuit
construction.

\paragraph{Terms}
We use the term \emph{throughput} to mean an amount of traffic that an
application or a segment of the network stack (e.g., TCP) has been measured to have forwarded (i.e. received and then sent). We use the term \emph{capacity} to
mean the maximum throughput that an application or network segment can handle.
Thus a \emph{Tor throughput} is an amount of traffic that a Tor process has been
measured to have forwarded, potentially as an estimate of \emph{Tor capacity}.
Tor throughput includes cell payloads and headers but excludes
TCP, IP, and other network packet headers. Finally, \emph{Tor ground truth} is
an estimate of Tor capacity experimentally determined by sending load from
increasing numbers of simulated clients and measuring them at the relay. Tor
ground truth measurements are accurate but expensive and require trust in the
relay.

\Section{TorFlow Analysis\pagebudget{2}}
\label{sec:torflow}

A primary goal of Tor's load balancing system is to drive more user
traffic to Tor relays that can better support it. Fundamentally, Tor
attempts to spread user load among relays according to two
relay characteristics: (i) relay capacity, i.e., the maximum rate at which
a relay can forward Tor traffic; and (ii) relay
performance, i.e., the current speed of fixed-size download. As
explained in \S~\ref{sec:background}, TorFlow uses relays' advertised
bandwidths as an estimate of relay capacity, and its own client
measurements to estimate relay performance.
Because TorFlow uses relay capacities as the basis for the load-balancing
weights it produces, accurate relay capacity estimates are essential to
load balancing and, ultimately, Tor network performance and scalability.

A relay's capacity estimate is derived from a heuristic measure of
unknown accuracy, i.e., its observed bandwidth (see
\S~\ref{sec:background}).
The observed bandwidth is likely inaccurate in many realistic cases: (i) a \textit{new}
relay will not have forwarded any traffic and thus will be estimated to have a
low capacity regardless of its available resources; (ii) a relay that clients
use \textit{inconsistently} may not sustain a high throughput long enough to result in an
accurate capacity estimate; (iii) a relay that clients \textit{underutilize} will
underestimate its capacity; and (iv) a relay with
co-resident processes that consume bandwidth \textit{inconsistently}
may overestimate its capacity.

To better understand the accuracy of Tor's capacity-estimation heuristic and its
effect on load balancing, we analyze publicly available Tor metrics data over
time. Relays' capacity estimates are published in their server
descriptors~\cite[\S~2.1.1]{tor-protocol-spec}, and load-balancing weights are
published in network consensus files~\cite[\S~3.4.1]{tor-dir-spec}. The Tor
Project has collected these documents for over a decade, and it publishes
monthly archives~\cite{tormetrics}. We use 11 years of such data from the
period 2008-08-01 to 2019-07-31, and we analyze the error in both
capacity estimation and load balancing.

\Subsection{Capacity Estimation Analysis}
\vspace{-2mm}

We compute the inaccuracy of relay capacity
estimates across relays and over time. In our analysis, we suppose that a
relay's true capacity does not often change (it usually runs on the same machine
with the same network interface card and access link) and that a relay's
observed bandwidth is not higher than its true capacity (a relay's capacity is not usually limited by co-resident processes).

\paragraph{Relay Capacity Accuracy}
We observe that advertised bandwidths exhibit high variance over time, which
suggests that they often underestimate true capacity (see Appendix~\ref{sec:torflowvariation}). To
quantify this error, we use a relay's maximum advertised bandwidth over a given
time period as a proxy for its true capacity. Comparing advertised bandwidths to
this maximum should yield a conservative error estimate, as the maximum should
only be lower than the true capacity, assuming the true capacity does not vary
during the given period.

\begin{figure}[t]
	\centering
	\includegraphics[width=0.8\columnwidth]{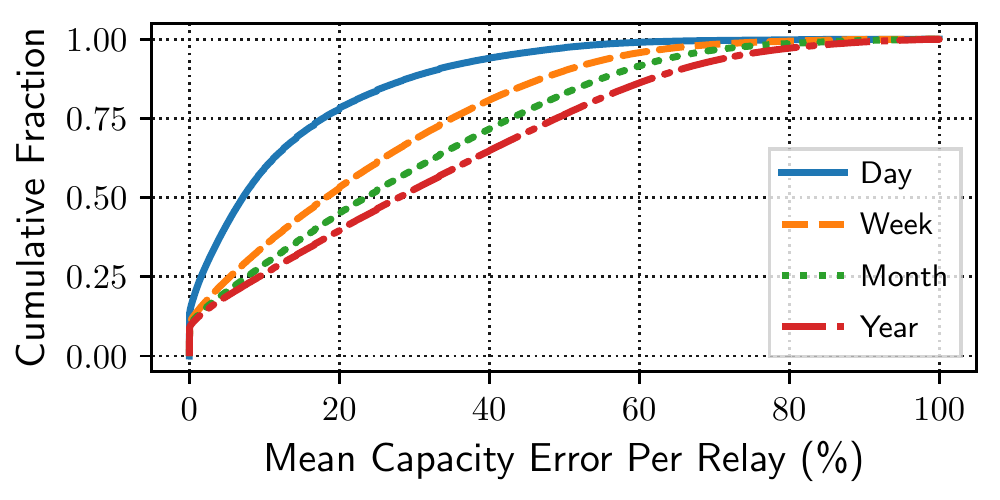}
	\vspace{-5mm}
	\caption{
	Relative error in relay capacity, computed using 11~years of archived Tor metrics data~\cite{tormetrics}.
	}
	\label{fig:relaycaperr}
	\vspace{-5mm}
\end{figure}

Let $A(r, t)$ be the advertised bandwidth of relay $r$ at time $t$, and let
$A(r,t,p)$ be the multiset of advertised bandwidths published during the period
of length $p$ preceding time $t$.
Varying the length of time $p$ allows us to examine error over different timescales.
We thus estimate the true capacity of relay $r$ at time
$t$ using the maximum bandwidth during the period of some length $p$ preceding
$t$:
\vspace{-2mm}
\begin{equation} \label{eq:capacity}
C(r, t, p) = \max\left(A(r, t, p)\right).
\vspace{-2mm}
\end{equation}
We can then assess the \textit{relay capacity error} for relay $r$ at
time $t$ as the fraction by which its advertised bandwidth underestimates the
maximum observed in the previous $p$ time:
\vspace{-2mm}
\begin{equation} \label{eq:relaycaperr}
\textrm{RCE}(r, t, p) = 1 - A(r, t)/C(r, t, p).
\vspace{-2mm}
\end{equation}
We summarize these errors by computing the mean of
$\textrm{RCE}(r, t, p)$ over the times $t$ on the hours between 2009-08-01 and
2019-07-31. We plot the distribution of these means over all
relays $r$ in Figure~\ref{fig:relaycaperr} for various values of $p$.

From the results in Figure~\ref{fig:relaycaperr}, we observe that larger errors
are estimated when the true capacities are based on longer time periods: the
median of the mean capacity error is 28\% when true capacities are computed
using 1 year of reports, compared to 7\% when they are computed using only 1 day
of reports. A plausible explanation for this observation is that relays are
typically underutilized and experience random load fluctuations, and so the
longer a relay is observed the more likely it is to receive traffic at or close
to its true capacity. We also find that over 85\% of relays have non-zero
capacity error, and for 25\% of relays the capacity error is 18\% or greater for
$p$~=~1~day and 49\% or greater for $p$~=~1~year. Overall, our results indicate
significant underestimation of true capacities.

\paragraph{Network Capacity Accuracy}
Although Figure~\ref{fig:relaycaperr} shows
non-trivial error in the \textit{relay} advertised
bandwidths, it does not necessarily indicate that the \textit{network}
suffers from inaccuracy as a whole. For example, it could be the case that
relays with highly erroneous advertised bandwidths are
slow relays that do not carry much user traffic. Therefore, we also
explore a notion of network accuracy. We compute the \textit{network
capacity error} at time $t$ by summing the advertised
bandwidths and true capacities from all relays and then
calculating the fraction of total network underestimation:
\vspace{-2mm}
\begin{equation} \label{eq:networkcaperr}
\textrm{NCE}(t, p) = 1- \sum_{r} A(r, t, p) / \sum_{r} C(r, t, p).
\vspace{-2mm}
\end{equation}
The network capacity error gives us an understanding of the total
fraction of Tor's capacity that is being underestimated, as opposed to
the fraction of relays with underestimation.

\begin{figure}[t]
	\centering
	\includegraphics[width=0.8\columnwidth]{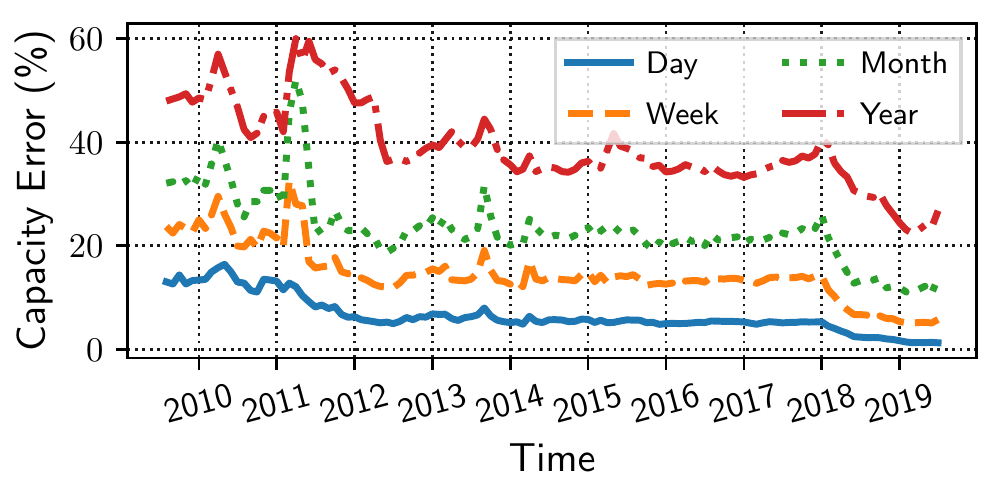}
	\vspace{-5mm}
	\caption{
	Network capacity error over time, computed using 11~years of archived Tor metrics data~\cite{tormetrics}.
	}
	\label{fig:networkcaperr}
	\vspace{-3mm}
\end{figure}

We show Tor's network capacity errors over time in
Figure~\ref{fig:networkcaperr}. In the median hour between 2009-08-01 and
2019-07-31, we find a network capacity underestimate of 5\% when using the
preceding day to determine true capacities, 14\% when using the preceding week,
22\% when using the preceding month, and 36\% when using the preceding year. As
with relay capacity errors, we see larger errors when we determine the true
capacity based on longer periods. The largest network capacity error we
discovered was 60\% (for $p$~=~1~year). These results provide evidence that Tor
significantly and consistently underestimates its total capacity. We discuss
additional conclusions in \S~\ref{sec:implications}.

\Subsection{Load Balancing Analysis}

Our previous results quantify the inaccuracy present in the
relays' advertised bandwidths. A main reason such inaccuracy is a problem is
that it affects the consensus weights used by clients to balance load across
relays. We analyze these consensus weights over time to better understand
the accuracy in the load balancing system.

\begin{figure}[t]
	\centering
	\includegraphics[width=0.8\columnwidth]{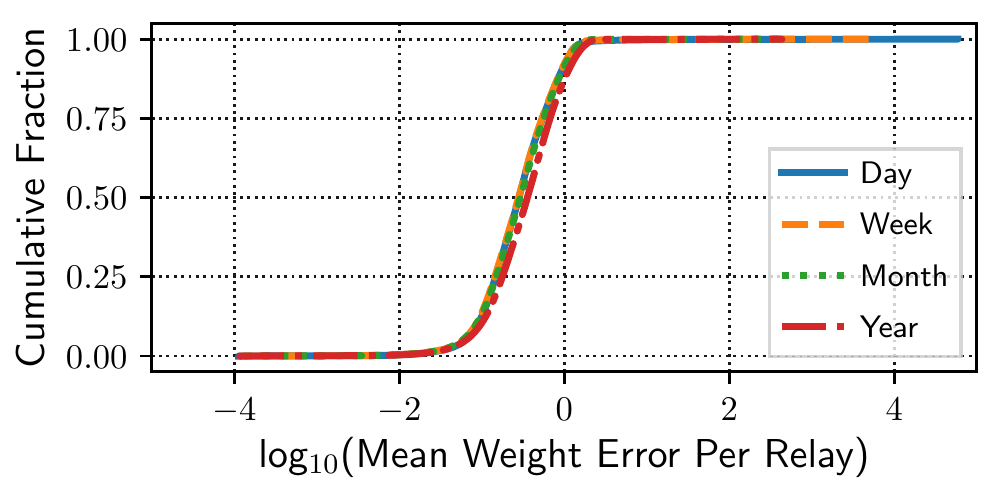}
	\vspace{-5mm}
	\caption{
	Relative error in relay weights, computed using 11 years of archived Tor metrics data~\cite{tormetrics}.
	}
	\label{fig:relayweighterr}
	\vspace{-3mm}
\end{figure}

\paragraph{Relay Weight Accuracy}
The probability that a relay is selected in a circuit is roughly its normalized
consensus weight, that is, its fraction of the total weight assigned to
all relays. Let $W(r, t)$ represent this value for relay $r$ at time $t$, and
let $W(r,t,p)$ be the multiset of these values over the consensuses during the
period of length $p$ preceding time $t$.
Ideally, a relay's normalized consensus weight would equal its normalized
capacity, that is, its fraction of the total capacity. Let $\overline{C}(r,
t,p)$ be the normalized capacity of relay $r$ at time $t$:
\vspace{-1mm}
\begin{equation} \label{eq:relaycapnormed}
\overline{C}(r, t, p) = C(r, t, p) / \sum_{s} C(s, t, p).
\vspace{-2mm}
\end{equation}
At any time $t$, we can consider any deviation of the normalized
consensus weight $W(r, t)$ from the normalized capacity
$\overline{C}(r, t,p)$ as error. We can then quantify this
\textit{relay weight error} by computing the ratio of these values:
\vspace{-2mm}
\begin{equation} \label{eq:relayweighterr}
\textrm{RWE}(r, t, p) = W(r, t) / \overline{C}(r, t, p).
\vspace{-2mm}
\end{equation}%
We then collapse the results to a single value per relay by computing
the mean over all $t$ starting from $t$~=~2009-08-01. Notice that it
is possible that the normalized consensus weight is less than the
normalized capacity, in which case $\textrm{RWE}(r, t, p) \in [0,1)$, and
that the normalized consensus weight is greater than the normalized
capacity, in which case $\textrm{RWE}(r, t, p) > 1$. Therefore, to better
visualize the results, we plot in Figure~\ref{fig:relayweighterr} the
distribution of the per-relay means of $\textrm{RWE}(r, t, p)$ (over all $t$)
by taking the $log_{10}$ of the means.
As a result, x-axis
values less than 0 indicate that relays are under-weighted compared to
their capacity, and x-axis values greater than 0 indicate that relays
are over-weighted compared to their capacity (by a factor of 10 for
each unit). The results show that more than 85\% of relays are
under-weighted compared to what we would expect based on their
capacities, while few relays (at $x$~=~0) are ideally weighted.
Since consensus weight is zero-sum, it must be the case that the
disproportionately small number of relays that are over-weighted
account for a disproportionately large amount of total user load;
we account for this in the following metric.

\begin{figure}[t]
	\centering
	\includegraphics[width=0.8\columnwidth]{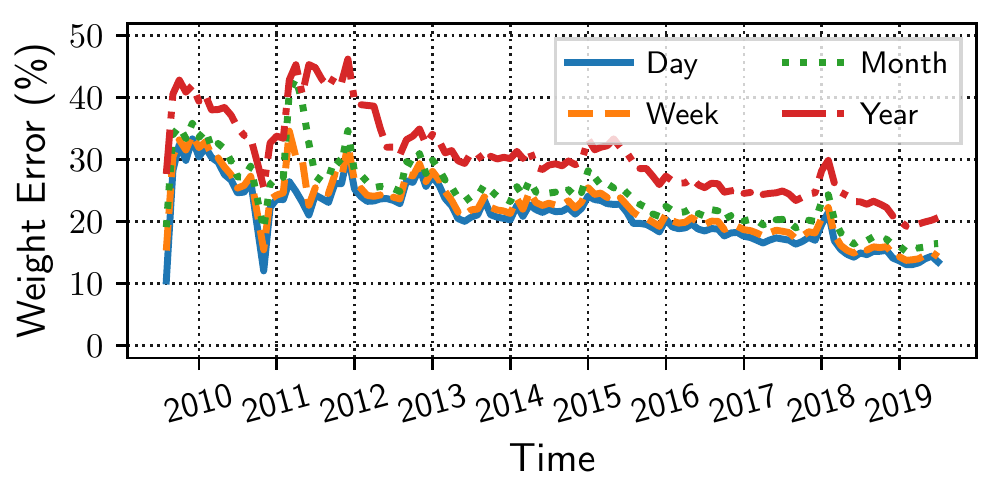}
	\vspace{-5mm}
	\caption{
	Network weight error over time, computed using 11 years of archived Tor metrics data~\cite{tormetrics}.
	}
	\label{fig:networkweighterr}
	\vspace{-3mm}
\end{figure}

\paragraph{Network Weight Accuracy}
To get a better sense of how weight errors affect the network overall
rather than individual relays, we compute the \textit{network weight
error} as the \textit{total variation distance} between the normalized
consensus weight and the normalized capacity:
\vspace{-2mm}
\begin{equation} \label{eq:networkweighterr}
\textrm{NWE}(t, p) = \frac{1}{2} \sum_{r} \mid W(r, t) - \overline{C}(r, t, p) \mid.
\vspace{-2mm}
\end{equation}
This better represents error in Tor's load balancing system overall,
since subtracting the normalized values (rather than dividing them as we
did in Equation~\ref{eq:relayweighterr}) means that relays will
contribute to the network error proportional to the amount of traffic
they carry.
Figure~\ref{fig:networkweighterr} shows the network weight error
$\textrm{NWE}(t, p)$ over time starting from $t$~=~2009-08-01 for various
values of $p$. We again observe that error increases as our capacity
estimates are based on data from longer periods. However, the
difference in error over greater values of $p$ is much less
pronounced: the network weight error is 21\%, 22\%, 24\%, and 30\% in
the medians when using normalized capacities from the preceding day,
week, month, and year, respectively. Our results over the latest year
of data (2019) show a 15-25\% error in load balancing weights,
indicating that Tor will benefit from improvements to their load
balancing system.

\Subsection{Conclusions and Observations} \label{sec:implications}

Our analysis shows consistent error in both network capacity and
network weights over time. We make two observations from our results.
First, we observe that we get a significantly higher estimate of true
capacity when aggregating more advertised bandwidths into the
estimate, and we generally find significant under-weighting of relays
relative to their capacity. Second, we observe that both capacity and
weight error has decreased in recent years (2018--2019) compared to
early years (2010--2011). We hypothesize that these observations
result from the under-utilization of relays: a relay that is not fully
utilized will report an advertised bandwidth that is below its true
capacity. We suspect that the error has decreased in more recent years
because Tor's increase in relay bandwidth resources has outpaced its
increase in user load, and therefore even after aggregating advertised
bandwidths over a year, our estimate of true capacity is still an
under-estimate (leading us to compute a lower error). We further explore this
hypothesis through a Tor network measurement experiment.

\Subsection{Relay Speed Test Experiment} \label{sec:speedtest}

To test our hypothesis that advertised bandwidths reported by relays
under-estimate their true capacity, we designed a relay speed test
experiment in which we flood each Tor relay with traffic for 20
seconds. The additional traffic that we transfer through the relays
will cause them to produce better estimates of their true capacities,
which they will then report in their server descriptors (see
\S~\ref{sec:bwbackground}).

\paragraph{Setup}
We added 487 lines of code to Tor v0.3.5.7 to support a new
\texttt{SPEEDTEST} cell that, when sent from a client to a supporting
relay, would simply be forwarded back to the client on the same
circuit. We also added client controller commands to start and stop a
speed test with a particular target relay and to monitor bandwidth
information during each test. We created 10 measurement ``teams'' that
each consisted of a client and a relay running our modified Tor, and
we ran a master python script that would keep track of the relays
available in the network consensus over time. The master script
iterated through the online relays one at a time, directing the speed
test client in each team to create a circuit through the target relay
to the speed test relay in that team. Once all teams' circuits were
built, the speed test clients and relays sent and forwarded
\texttt{SPEEDTEST} cells as fast as possible for 20 seconds, taking
care not to overflow circuit queue length limits. This resulted in 20
new bidirectional, high volume sockets on each target relay, who
forwarded the traffic as they would on any other circuit. We ran all
speed test clients and relays on the same dedicated machine with 32
GiB of RAM, 8 CPU cores, and a 1 Gbit/s network link.

\paragraph{Ethics}
We designed our experiment to minimize Tor network relay overhead.
We submitted our experimental design and plans to the Tor Research
Safety
Board\footnote{\url{https://research.torproject.org/safetyboard}} for
feedback. We received encouraging feedback and a ``no objections''
decision. We also explained our plans to the Tor community through a
post to the public \textit{tor-relays} mailing list. We gave
instructions on how to opt out and allowed one week to collect
feedback. Finally, we served a web page containing a link to the
mailing list post on the IP addresses used in the experiment.

\begin{figure}
	\centering
	\includegraphics[width=0.8\columnwidth]{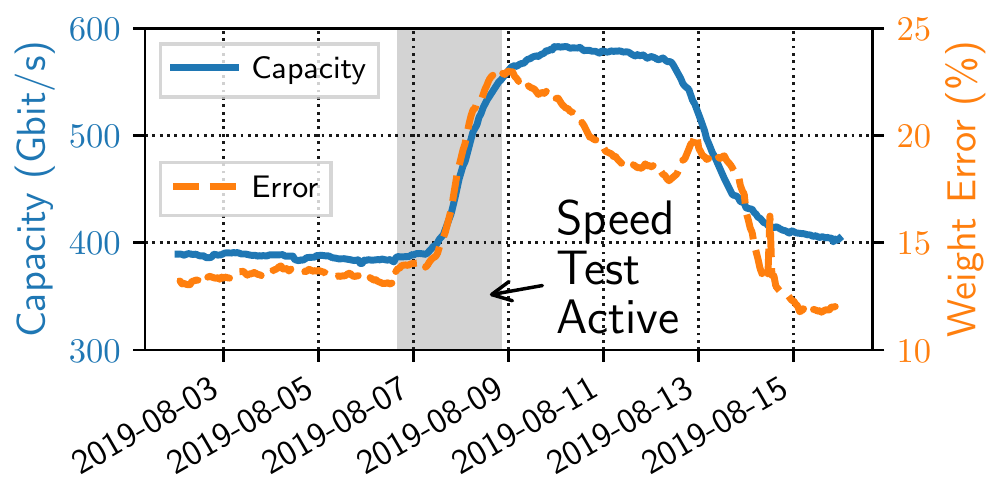}
	\vspace{-3mm}
	\caption{
	Our relay speed test discovered $\approx$200 Gbit/s of excess capacity ($\approx$50\%), and the network weight error (Equation~\ref{eq:networkweighterr}) increased by between 5\% and 10\% due to more accurate capacity estimates.
	}
	\label{fig:speedtest}
	\vspace{-3mm}
\end{figure}

\paragraph{Results}
Our speed test experiment ran for just over 2 days (51 hours) starting
on 2019-08-06, as shown in the shaded region in
Figure~\ref{fig:speedtest}. During this time we successfully measured
4,867 relays and we observed timeouts for 2,132 relays. We plot in
Figure~\ref{fig:speedtest} the estimated capacity of the network,
i.e., the sum of advertised bandwidths for all online relays over
time. We find that the estimated network capacity increases by about
200 Gbit/s (about 50\%) after our speed tests push relays into
reporting higher observed bandwidths, and that 10\% of relays reported
a change in observed bandwidth of 140 Mbit/s or more. Note that the
delay in the increase and decrease in capacity relative to our
experiment is caused by (i)~the 18 hour server descriptor publishing
interval, and (ii)~the observed bandwidth algorithm which stores
history for each of the last 5 days. We also plot in
Figure~\ref{fig:speedtest} the network weight error as defined in
Equation~\ref{eq:networkweighterr}, which represents the overall
effect on load balancing. We find that weight error increased by
between 5\% and 10\% as a result of more accurate capacity estimates,
to a maximum of 23\% during the speed test. Then, we observe a
decrease in weight error immediately following our experiment, which
we believe is a result of TorFlow using the new information to correct
consensus weights. From these results, we conclude that Tor
significantly under-estimates its available network capacity and that
better estimates of capacity would reduce error in load balancing.

\Section{\tool{} Design \pagebudget{2}} \label{sec:design}

We now present the design for \tool{}, a system to measure the capacity of Tor
relays. The key technique behind \tool{} is to actively measure the full
capacity of Tor relays using multiple measurement hosts. This approach improves security over prior approaches,
as relays must demonstrate their true capacity, a process that cannot be faked. It also improves accuracy, as the measurement does not depend
on background traffic or on other relays.

\paragraph{Setup}
\tool{} uses a \emph{measurement team} to perform relay capacity measurements.
The measurement team consists of a set of \emph{measurers} running on hosts
whose resources are dedicated to the measurement process. The measurers will
cooperatively measure relays, and so the primary requirement for measurers is
that they collectively have sufficient network capacity to measure all Tor
relays.
A team is considered to have sufficient capacity if the sum of capacities over
all measurers is at least some constant factor $f$ (see \S~\ref{sec:deployment})
times the highest Tor-relaying capacity among relays. \tool{} is designed to
achieve accurate measurement given sufficient network capacity, regardless of
network latency.

The measurement team is coordinated by a BWAuth, who determines the measurement
\emph{schedule} and aggregates the results. A measurement schedule is created for each measurement
\emph{period}, which divides time into constant-length intervals. Multiple BWAuths,
each with its own measurement team, independently run \tool{}. Each BWAuth separately
measures each relay during a period.

\paragraph{Trust and Diversity}
As in Tor currently, each DirAuth
chooses to trust some BWAuth, and the DirAuths place the median of their
measurements in the consensus. Thus the trust assumption in FlashFlow is that a majority of
DirAuths trust BWAuths (and their associated teams) that are honest. In
the simple case that each DirAuth trusts a different BWAuth, \tool{}
requires an honest majority among the BWAuths and their teams. Moreover, \tool{}
will be more accurate if there is a diversity of network locations across measurement teams that
reflects the diversity of Tor clients and relays. Such diversity will mitigate unrepresentative
measurements resulting from unusually high or low network capacity between a relay and a measurement
team (e.g., due to existing in the same data center).

\Subsection{Performing a Measurement} \label{subsec:single-measurement}
A BWAuth initiates a single measurement by creating an authenticated connection to
each measurer and to the target relay. Authentication is performed using the public key of the
BWAuth, which we assume is distributed in the Tor network consensus. The BWAuth sends the target
the public keys of each measurer involved in the measurement. While connected, the measurers accept
instructions from the BWAuth, and the relay accepts authenticated measurement connections
from the measurers indicated by the BWAuth. The relay will only accept connections from a given
BWAuth and its team once per measurement period.

The BWAuth will divide the total resources needed for the measurement across its $m$ measurers
$M_1,\ldots, M_m$. The BWAuth allocates a quantity $a_i$ of the measurement capacity of $M_i$ to the
measurement (see \S~\ref{subsec:measuring-relay} for choosing $a_i$), where $a_i=0$ is possible
and indicates that $M_i$ does not participate in the measurement. For each $M_i$, a modified Tor
process is started on each CPU core without an existing measurement process (and always at least
one). The measurement-traffic rate of the $k_i$ processes thus started is limited by setting the
\textsf{BandwidthRate} parameter of each modified Tor process on the measurer to $a_i/k_i$. A
constant total number of TCP sockets $s$ is used across all measurers (see Appendix~\ref{sec:num-socks}
for setting $s$), and each $M_i$ uses an even share $s/m$ of them, with each measuring process at
$M_i$ using $s/(mk_i)$ of the sockets.

Each measuring process creates one TLS connection with the target relay for each of its allocated
sockets. Over each such connection, a special measurement circuit is constructed using a new type of
circuit-creation cell. A key exchange is performed, but the circuit will not be extended further.
All cells received on the circuit by the target relay will be decrypted and then returned to the
measurer. The target relay schedules cells on measurement circuits using a separate cell scheduler
to ensure high throughput even with fewer sockets than typical for a Tor relay (the existing
scheduler~\cite{kist-tops2018} is designed for priority scheduling across many
sockets~\cite{trac29427}). Moreover, the target relay enforces a maximum ratio $r$ between cells
sent by the normal scheduler and those overall, and it attempts to send as
much normal traffic subject to this maximum. This design provides an accurate measurement while
ensuring that normal traffic continues to be relayed.

A relay is measured by a BWAuth during a measurement \emph{slot}. During this time, each measuring
process sends measurement cells filled with random
bytes over the measurement circuit. The process sends such cells as fast as possible. The target
relay decrypts those cells using the circuit key, and then
returns them on the circuit. Note that both the measurer and the target perform TLS
encryption and decryption, but the target alone performs Tor's cell decryption.
This design minimizes
the computational load of the measurer while replicating the cryptographic operations that the
target would perform on normal traffic, which is needed to get an accurate estimate of its forwarding capacity.
To ensure that the target is correctly decrypting and
forwarding cells, the measurer records the contents of each cell sent with probability $p$ (e.g.,
$p=10^{-5}$) and checks that the returned content of such cells is correct, reporting failure from
the measurement if not.

A measurement slot lasts a constant number of seconds $t$ (see
\S~\ref{sec:deployment}). The BWAuth can end the measurement in this slot early
due to a failure reported by a measurer. During the measurement slot, the BWAuth
receives from the $i$th measuring process the number of measurement bytes
$x_j^i$ that were relayed by the target to the process in the $j$th second. The
BWAuth also receives from the target the number of normal traffic bytes $y_j$
that the target relayed in the $j$th second. At the end of the measurement, the
BWAuth computes per-second sums of measurement traffic: $x_j = \sum_{i=1}^m
x_j^i$. It limits the per-second normal traffic to the largest value that is
consistent with the measurement traffic and the traffic ratio $r$:
$\overline{y}_j = \min\left(y_j, x_j r/(1-r) \right)$. The BWAuth computes a
per-second estimate $z_j = x_j + \overline{y}_j$ of total bytes relayed by the
target, and then it sets its capacity estimate to the median: $z =
\textrm{median}\left(z_1,\ldots, z_t\right)$. Incorporating the normal traffic
results in better capacity estimates, and enforcing the expected ratio limits
how much a malicious relay can increase its capacity estimate by reporting more
normal traffic than it actually relayed.

\Subsection{Measuring a Relay} \label{subsec:measuring-relay}
\vspace{-2mm}
Measuring a relay potentially involves a sequence of measurements because the measurer capacity
required for an accurate measurement is unknown. Instead of using the maximum amount of measurer
capacity for each relay, we instead use informed guesses about relays' capacities and
allocate only the measurer capacity needed for those guesses. If the measurement indicates that the
allocated capacity was sufficient for a given target, then we conclude the measurement process.
Otherwise, we perform another measurement of the target with a higher guess and more measurer
capacity. This process reduces the total amount of measurer capacity used to measure the
entire network.

\paragraph{Measuring Measurers}
To allocate measurer capacity, we first need to estimate the network forwarding capacity of the
measurers. Measuring measurers is easier than measuring relays because (i) we only need a lower
bound on the measurement capacity, as an underestimate will only affect the speed of the measurement
process and not its accuracy; and (ii) we only need to measure the speed at which network traffic can
be simultaneously sent and received, as the measurer doesn't relay bytes through Tor. Therefore, to
estimate the network forwarding capacity of a measurer, the BWAuth instructs it to use
iPerf~\cite{iperf} to exchange bidirectional traffic with each other measurer on the team
concurrently. This measurement uses UDP to eliminate the effects of TCP congestion control that are
unlikely to affect the measurement of all relays. This measurement need only be performed when a new
measurer is added to the team or when the BWAuth expects the capacity of a measurer to have changed.
A 60-second measurement is performed, and the capacity estimate is the median of the per-second
speeds reported by iPerf.

\paragraph{Measuring Old Relays}
When measuring an \emph{old} relay, that is, one that has an existing capacity estimate $z_0$, we
simply use $z_0$ as a guess for its current capacity. The BWAuth needs to allocate $f\cdot z_0$
total capacity across the measurers, where $f$ is an excess allocation factor. Let $c_i$ denote the
network capacity of measurer $M_i$. The BWAuth can allocate to this measurement any amount $a_i$ of
the capacity of $M_i$ subject to $0\le a_i\le c_i$ and $\sum_i a_i = f\cdot z_0$. We greedily
allocate capacity by repeatedly assigning the measurer with the most residual capacity to use
all its remaining capacity or as much as is needed to reach $f\cdot z_0$.

The allocation factor $f$ is defined so that the measurement has a high probability of being
accurate and conclusive. It depends on a multiple $m$ that is just large enough so that, for error
parameters $\epsilon_1, \epsilon_2\ge 0$, if a relay with true capacity $x$ is measured using at
least $m x$ measurer capacity, then the capacity estimate $z$ is almost certainly greater than
$(1-\epsilon_1)x$ and less than $(1+\epsilon_2)x$. The value for $m$ is determined experimentally (see Appendix~\ref{sec:mult}). In addition to $m$, $f$ includes a factor $(1+\epsilon_2)/(1-\epsilon_1)$ to
ensure that $z$ cannot result from values $x'>x$ for which the measurement errors may be larger. The
excess allocation factor is thus $f = m(1+\epsilon_2)/(1-\epsilon_1)$.

Using the capacity allocations, the team performs a measurement and obtains a capacity value $z$.
This value is taken as the new estimate if it is small enough relative to the total measuring
capacity that it could only result from a true relay capacity close to $z$. Specifically, $z$ is
the new capacity estimate if $z < \sum_i a_i (1-\epsilon_1)/m$. When this is true, the true relay
capacity $x$ must be greater than $z/(1+\epsilon_2)$ and less than $z/(1-\epsilon_1)$, which implies
that the estimate is accurate, i.e., that $z\in ((1-\epsilon_1)x, (1+\epsilon_2)x)$.
If the capacity estimate $z$ is not sufficiently small, then the relay must be measured again using
a higher total measurer capacity. In this case, we set $z_0=\max\left(z, 2z_0\right)$, which
ensures the allocated capacity will at least double, and we repeat the measurement with the
updated capacity estimate $z_0$.

Observe that if the original estimate $z_0$ is the true capacity, then the measurement process will
almost certainly conclude after one measurement. This is true because the measuring capacity was
chosen to be large enough that $z < (1+\epsilon_2)z_0$ with high probability, and when that is
true the condition to use the $z$ as the new capacity estimate is satisfied:
\begin{equation*}
z < z_0 (1+\epsilon_2) = z_0 f (1-\epsilon_1) / m = \sum_i a_i (1-\epsilon_1)/m.
\end{equation*}

\paragraph{Measuring New Relays}
When measuring a \emph{new} relay (i.e., one without a capacity estimate), we initially guess
the capacity based on the capacity distribution of existing relays. New relays either have never
been seen before or were last measured so long ago (e.g., a month) that their capacity measurements
are no longer considered reliable estimates. For such relays, we use as a capacity estimate $z_0$
the 75th percentile measured capacity among Tor relays over the past month. When this value is
sufficiently smaller than the maximum capacity measurable, this allows us to devote less measurer
capacity to the measurement. We then expect that one measurement will be sufficient for 75\% of new
relays. Given this estimate, the measurement proceeds the same as with old relays, where again if
the resulting measurement $z$ is too high relative to the allocated capacity, the relay is
scheduled for another measurement with an updated estimate $z_0=\max\left(z, 2z_0\right)$.

\Subsection{Measuring the Network} \label{subsec:measuring-network}
To measure all relays in the network, the BWAuths periodically determine the
measurement schedule. The schedule determines when and by whom a relay should be measured. We
assume that the BWAuths have sufficiently synchronized clocks to facilitate
coordinating their schedules. A measurement schedule is created for each measurement period, the
length $p$ of which determines how often a relay is measured. We use a measurement period of $p=24$
hours.

To help avoid active denial-of-service attacks on targeted relays, the measurement schedule is
randomized and known only to the BWAuths. Before the next measurement period starts,
the BWAuths collectively generate a random seed (e.g., using Tor's secure-randomness
protocol~\cite{tor-randomness-spec}). Each BWAuth can then locally determine the shared
schedule using pseudorandom bits extracted from that seed. The algorithm to create the schedule
considers each measurement period to be divided into a sequence of $t$-second measurement slots.
For each old relay, slots for each BWAuth to measure it are selected uniformly at
random without replacement from all slots in the period that have sufficient unallocated measurement
capacity to accommodate the measurement. When a new relay appears, it is measured
separately by each BWAuth in the first slots with sufficient unallocated capacity. Note
that this design ensures that old relays will continue to be measured, with new relays given
secondary priority in the order they arrive.

\Section{Security Analysis \pagebudget{1.5}}\label{sec:security}

\paragraph{Properties}
\tool{} is designed to be secure against an adversary that attempts to cause
incorrect measurements. For the specific application of load balancing, we are
particularly focused on preventing malicious relays from obtaining incorrectly
large capacity estimates and honest relays from obtaining incorrectly small
estimates. The threat model includes an adversary that runs malicious relays,
malicious clients, some malicious BWAuths, and some malicious DirAuths. Honest
BWAuths are assumed to use honest measurement teams. We require that a majority
of the DirAuths trusts honest BWAuths.

\looseness-1
The \tool{} design requires a target relay to demonstrate its capacity in a way that cannot be
falsified. Thus, rather than depending on self-reports (as TorFlow does fundamentally), \tool{}
has measurers actually send and receive the same cells as normal Tor clients would. Moreover,
the sent cell contents are randomly generated and the received contents checked at random to ensure
that the target is properly receiving, decrypting, and returning the cells during the measurement.
A relay that forges responses (e.g., to skip decryption or to send early before receiving) is
detected with overwhelming probability when a response cell is checked due to the random contents,
and a response cell is checked with probability $p$. As a result, a malicious relay that forges $k$
responses has approximately a $(1-p)^{-k}$ chance of evading detection.

Relays are trusted to some extent to report the normal client traffic that is forwarded during
a measurement. However, that client traffic is supposed to be limited to most a fraction $r$ of
the total traffic, and during aggregation the BWAuth limits the reported normal
traffic to be at most $r$ times the total. An honest relay will enforce the
ratio, and so the aggregated measurement accurately takes into account both types of traffic.
A malicious relay could send no normal traffic but report the full amount, and it could thereby
inflate its capacity estimate by a factor $1/(1-r)$ above the truth.

Several features also prevent a relay from providing high capacity only while it is being measured.
Measurement by any given BWAuth is performed at a randomly selected slot in a
measurement period, and the randomness is known only to the BWAuths. Furthermore, the
relay is measured by multiple BWAuths at separate random times, and the median of the
estimates is used. For an adversary that does not control a BWAuth, an attempt to
provide high capacity only during a fraction $q<1/2$ of measurement slots will fail with probability
at least 0.5. More accurately, with $n$ BWAuths the probability is $\sum_{k=n/2}^n Pr[B(n,1-q)=k]$,
with $B(n,p)$ binomially distributed. Relays are notified of a measurement at its beginning, but
due to the shortness of the measurement slot (e.g. $t=30$ seconds), a malicious relay has little
time to adjust its capacity dynamically. The frequency with which relays are measured also forces
malicious relays to be able to consistently support their measured capacities. A relay is measured
once every period, and so even after a relay has been measured by a majority of BWAuths (which is
expected to take a majority of the period), it can only reduce its capacity until the next period.
The efficiency of \tool{} allows the measurement period to be relatively short (e.g. every 24 hours)
and thus gives little time for a malicious relay to act at a reduced capacity.

Another security benefit of a randomized measurement schedule is that it limits the opportunity for
malicious clients to perform a targeted denial-of-service attack. An adversary may try to do this
in order to reduce the measured capacity of certain honest relays, which would cause Tor's load
balancing to shift traffic away from them. However, assuming the adversary controls no BWAuth, the adversary cannot predict when an honest relay will be measured and must perform any
denial-of-service attack during most of the measurement slots in order to expect to affect the
median measurement.

Finally, we observe that it is difficult for an adversary to prevent relays from being measured by
flooding the network with new relays (i.e., a Sybil attack). Old relays are guaranteed to be measured
during a measurement period because they are scheduled first. New relays are given second priority,
and moreover they are served on a first-come, first-served basis, and so benign new relays are eventually
measured.

\paragraph{Limitations}
In some cases malicious relays may be able to cause \tool{} to obtain larger capacity estimates
than the relays could sustain in Tor. We argue that these limitations are shared by Tor's
existing system, TorFlow, and that \tool{}'s security and accuracy advantages make it a
significant improvement. Moreover, we suggest ways to improve \tool{} in the
future to mitigate these issues.

One limitation is that an adversary that has access to multiple IP addresses on the
same machine can surreptitiously run multiple relays on the machine simultaneously.
Tor only accepts two relays at the same IP address (a restriction that was instituted as a defense
against falsely obtaining a large total bandwidth weight~\cite{tor-proposal-109}).
\tool{} is likely to measure multiple relays on the same machine at separate times, and so each
relay would obtain a capacity estimate that is the capacity of the shared machine. Tor considers
this a Sybil attack, and it currently requests that each relay operator identifies all relays that
they run with the \textsf{MyFamily} option~\cite{tor-manual}. Moreover, Tor has made use of systems
designed to detect Sybils on its network~\cite{winter2016identifying}. Pairs of \textsf{MyFamily}
relays (or suspected Sybils) can be measured simultaneously with \tool{} to determine if they share
the same Tor capacity, and then the measured capacity averaged over the members of a connected set.
The current TorFlow system shares this issue, as the speed measurements are performed at
different times, and an adversary can detect when one of its relays is being
measured and reserve all capacity for the measurement
circuit~\cite{thill-thesis,peerflow-pets2017}.

Another limitation is that \tool{} measurements are so short that they might
measure the \emph{burst} speed of a host rather than its sustainable Tor
capacity. For some ISPs and hosting providers, higher burst capacities are
supported than are consistently achievable. This can be true as a matter of
practice, as a network shared by many hosts may occasionally be
underutilized, or as a matter of policy, as providers may institute price-based
limits on the speed of network traffic from a host. In the former case, if the
burst speed is due to variable congestion of shared resources, then we expect
the median of the separate and randomly scheduled measurements by different
BWAuths to produce good estimates of average performance. In the latter case, if
such limits are applied faster than half the length of our measurement slots
(e.g., in less than 15 seconds), then \tool{} should obtain a sustainable
capacity estimate. Moreover, we again observe that
this issue currently affects TorFlow, which performs relatively short downloads
of files (none larger than 64\,MiB).

We further note that \tool{} is designed to measure Tor capacity and not to detect if client
traffic is actually relayed. A malicious relay can send little to no real client
traffic while obtaining accurate capacity estimates from \tool{} by only sending traffic on
measurement circuits. This is an additional limitation shared with TorFlow, in which the
measurement circuits are easily detected~\cite{thill-thesis} and thus weights can be obtained
while denying all traffic except not used for measurement~\cite{peerflow-pets2017}. Such behavior
seems highly observable, however, and so we leave detecting such misbehavior as a future enhancement.

\newcommand{\errlow}{0.20}
\newcommand{\errhigh}{0.05}

\Section{Network Experiments\pagebudget{3}}\label{sec:deployment}

We measure and evaluate \tool's performance and accuracy with a set of network experiments.

\Subsection{Preliminary Setup and Analysis}

\paragraph{Internet Vantage Points}\label{sec:environment}
To perform realistic measurements on the Internet, we obtain hosts from a
set of geographically diverse network locations.
Table~\ref{tab:host-summary} summarizes the characteristics of our hosts
located in
Fremont, CA (US-SW),
Santa Rosa, CA (US-NW),
Washington, DC (US-E),
Bangalore, India (IN), and
Amsterdam, Netherlands (NL).

\begin{table}
\centering
\footnotesize
\begin{threeparttable}
\captionsetup{skip=0pt} %
\caption{Summary of the hosts used in Internet experiments}
\label{tab:host-summary}
\begin{tabular}{ r |@{\,\,\,\,}c@{\,\,\,\,}c@{\,\,\,\,}c@{\,\,\,\,}c@{\,\,\,\,}c }
\toprule
	          & \textbf{US-SW} & \textbf{US-NW} & \textbf{US-E}  & \textbf{IN}    & \textbf{NL}    \\ %
\midrule
	\textbf{Virtual}   & No     & Yes     & No     & Yes     & Yes     \\
	\textbf{Network Type$^\star$}  & D.C.    & D.C.    & Res.   & D.C.    & D.C.    \\
	\textbf{BW (claimed) (Gbit/s)}    & 1000     & 1000     & 1000     & N/A   & N/A   \\
	\textbf{BW (measured) (Gbit/s)}  & 954   & 946   & 941   & 1076  & 1611  \\
	\textbf{RTT to US-SW (ms)}       & 0     & 40    & 62    & 210   & 137   \\
	\textbf{CPU cores} & 8     & 8     & 12    & 2     & 2     \\
	\textbf{RAM (GiB)}       & 32    & 4     & 32    & 4     & 4     \\
\bottomrule
\end{tabular}
\begin{tablenotes}
\footnotesize
\item[$\star$] Network type is datacenter (D.C.) or residential (Res.)
\end{tablenotes}
\end{threeparttable}
\vspace{-3mm}
\end{table}

Because network bandwidth is an important factor that will affect our
experiments, and because the supported bandwidth was not advertised for all
hosts, we empirically estimate it using iPerf~\cite{iperf} (a network
performance measurement tool). We perform a set of experiments where for each
host we instruct all other hosts to perform a UDP iPerf measurement to it at the
same time for 60 seconds. We sum together the per-second results from each host
and present the median of the summed per-second results in the ``BW (measured)''
row in Table~\ref{tab:host-summary}. All three of the US hosts are clearly
limited to about 1 Gbit/s. IN and NL achieve higher throughput, despite their
hosting provider making no claims about their capacity.
We present additional pairwise TCP and UDP
measurement results in Appendix~\ref{sec:internettput}.

\paragraph{Tor Processing Limits}\label{sec:processlimits}
We evaluate Tor's processing limits to estimate
the throughput that a \tool deployment must support in order to
measure the fastest Tor relays. We set up a lab experiment that
attempts to maximize throughput while minimizing the effect of
limiting factors including network latency, congestion and flow
control algorithms in TCP and in Tor, the capacity of the underlying
network, and the number of Tor circuits and TCP sockets used during
the measurement. Over a 120-second measurement, we found that a Tor relay was
able to process traffic at a rate of 1.25 Gbit/s in the median while using 20 TCP sockets. We
confirmed that Tor reached 100\% CPU utilization during this
measurement, which is expected because Tor runs all of its cell
scheduling code in a single thread. We conclude that the Tor software
should not prevent us from measuring even the fastest Tor relay,
the claimed capacity of which was 998 Mbit/s in July 2019~\cite{tormetrics}.
\begin{conference}
See Appendix~\ref{sec:tors-limit} for more details about the setup of
and results from this experiment.
\end{conference}

Because we will use our US-SW host to run target relays in our Internet
experiments, we also establish ground-truth Tor capacity on it
by running an experiment similar to the one described above.
We run a relay on US-SW, and use the remaining machines to run Tor
processes that support the measurement of US-SW.
The target relay on US-SW achieves a maximum median throughput of 890
Mbit/s while consuming 95--100\% of a CPU core (again, due to Tor's
primarily single-threaded nature). We conclude that this is the
fastest we can expect Tor to forward traffic on US-SW.

\begin{full}
More details about the above experiments appear in
Appendix~\ref{sec:tors-limit}. We also provide results in
Appendix~\ref{sec:kernel-tuning} from additional experiments that further verify
that we can fully measure Tor relay capacity.
\end{full}

\paragraph{\tool Implementation and Setup}
We implement \tool{} as a 1,200-line patch to Tor v0.3.5.7 containing
measurer- and relay-side measurement support and a 1,300-line C/Rust
program that controls \tool{} measurers.
The experimental setup for the remainder of this section is as follows.
US-SW runs a single target Tor relay. Some combination of the remaining hosts
(US-NW, US-E, IN, and NL) measure the target relay.
We configure \tool with the following settings, which were determined
through a sequence of experiments detailed in
Appendix~\ref{sec:derive-params}: the number of measurement sockets $s=160$ (the
$s$ that maximizes throughput on the slowest host); the multiplier $m=2.25$ (the
smallest $m$ that yields sufficient accuracy); the measurement duration and
strategy is to take the median throughput achieved in $t=30$ seconds
(reasonable balance between time-to-result and accuracy); and error bounds of
$\epsilon_1=\errlow{}$ and $\epsilon_2=\errhigh{}$. We consider the effect of
kernel tuning on $s$ in Appendix~\ref{sec:kernel-tuning}.

\Subsection{Measurement Accuracy}\label{sec:accuracy}

We evaluate \tool's accuracy with and without client background
traffic.

\paragraph{Without Client Background Traffic}\label{sec:verify}
\begin{figure}
	\centering\includegraphics[width=0.4\textwidth]{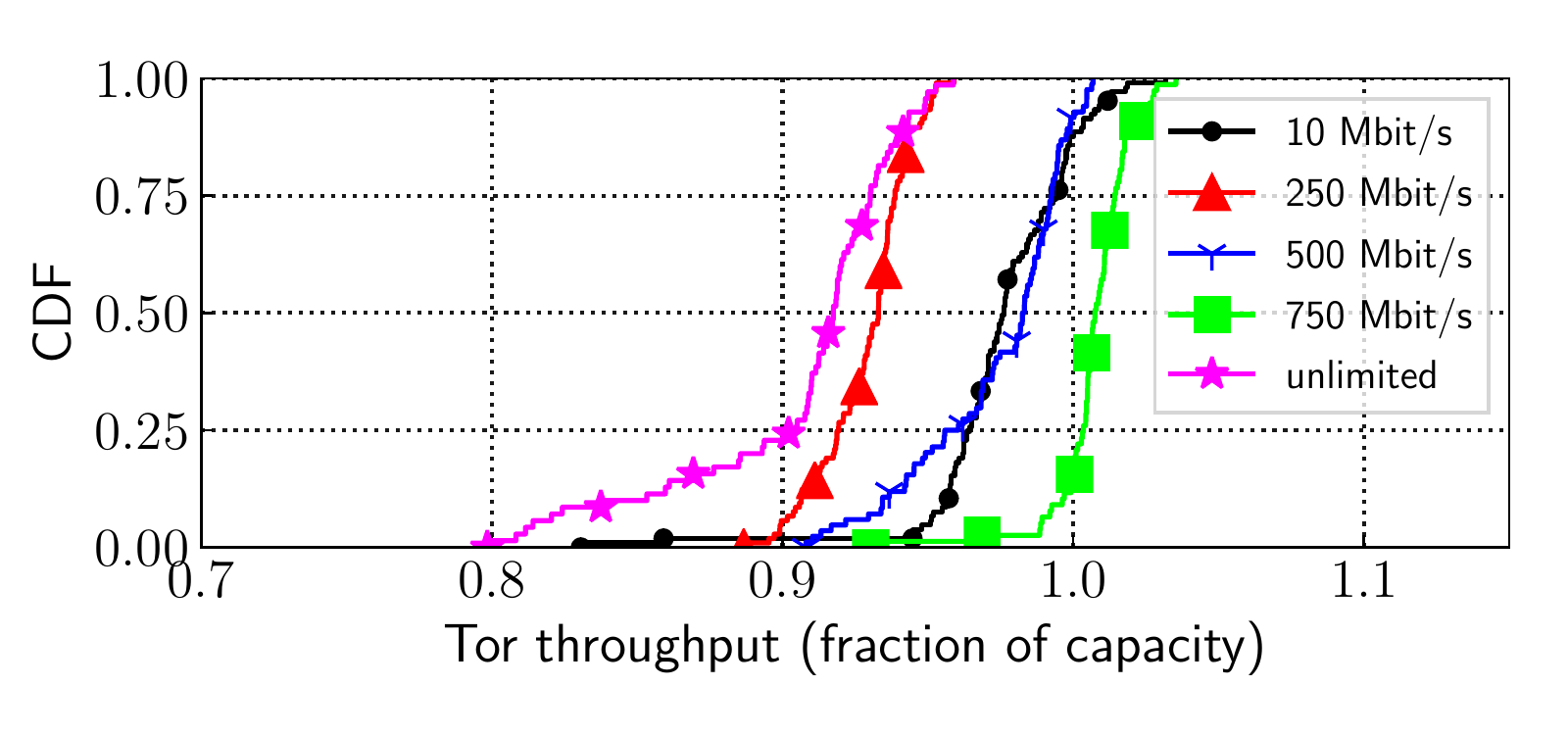}
	\vspace{-5mm}
	\caption{Evaluation of \tool{}'s accuracy from 24 hours worth of 30
	second experiments with multiplier $m=2.25$. CDFs are over the median
	per-second throughput measured by each team.}
	\label{fig:verify}
	\vspace{-5mm}
\end{figure}
We conduct a set of Internet
experiments in which we configure a target relay on US-SW and form
measurement teams from all possible unique subsets of the remaining
machines from Table~\ref{tab:host-summary}. We set throughput limits
of 10, 250, 500, 750, and unlimited Mbit/s on the target; for each
such limit we test how well all measurement teams can measure it,
where each measurer in each team is limited to its share of the factor
$f$ of measurer capacity that is necessary to measure the target using
$m=2.25$. Each such measurement (a team measuring a throughput-limited
relay) runs for 30 seconds and is repeated 7 times over the course of
24 hours. The result of each measurement is the median per-second
throughput over the 30 second period.

Figure~\ref{fig:verify} shows the accuracy of our measurements,
categorized by the Tor capacities at the target.
Across all configured
capacities, all but one experiment (99.8\%) produces results within
$\epsilon_1=\errlow{}$ and $\epsilon_2=\errhigh{}$. \tool{} measures within
11\% error (0.89--1.11 times capacity) in 95\% of experiments.
\begin{full}
We provide additional results in Appendix~\ref{sec:concurrent-measurements}
showing that measurements remain accurate when multiple relays are measured
concurrently, which would occur during a full-network Tor measurement by
\tool{}.
\end{full}

\paragraph{With Client Background Traffic}\label{sec:bg-traffic}
To evaluate \tool's ability to measure a relay with realistic client
background traffic, we run a Tor relay on US-SW and connect it to the
real Tor network.\footnote{In practice, we run the relay on a machine
with hardware identical to US-SW and in the same datacenter in order
to parallelize our experiments.} We run the relay for 60 days
before starting any \tool measurements so that it is
measured by the existing BWAuths, earns the Guard flag, and attracts
a significant amount of client traffic (50 Mbit/s on average).  The
relay is configured to limit its Tor throughput to 250 Mbit/s, and we
measure it with one \tool measurer running on NL. Before, during,
and after each experiment we record per-second Tor throughput events
from the relay that include the total amount of traffic it is
forwarding, while during the experiment \tool{} reports for each
second both the amount of measurement traffic that the relay is
fowarding as well as the amount of background traffic that it
claims to have forwarded.

Recall from \S~\ref{subsec:single-measurement} that a Tor relay
enforces a maximum ratio $r$ between its regular background traffic
and its measurement traffic during a measurement: a higher $r$ allows
for more regular traffic and minimizes the effect that the measurement
process has on the Tor clients, while a lower $r$ minimizes the
advantage a malicious relay has when lying. We evaluate \tool's
accuracy under multiple $r$ ratios; we present the results from
$r=0.1$ while noting that the results from the other tested values of
$r$ were similar.

\begin{figure}[t]
	\centering
    \captionsetup{skip=0pt} %
	\includegraphics[width=0.4\textwidth]{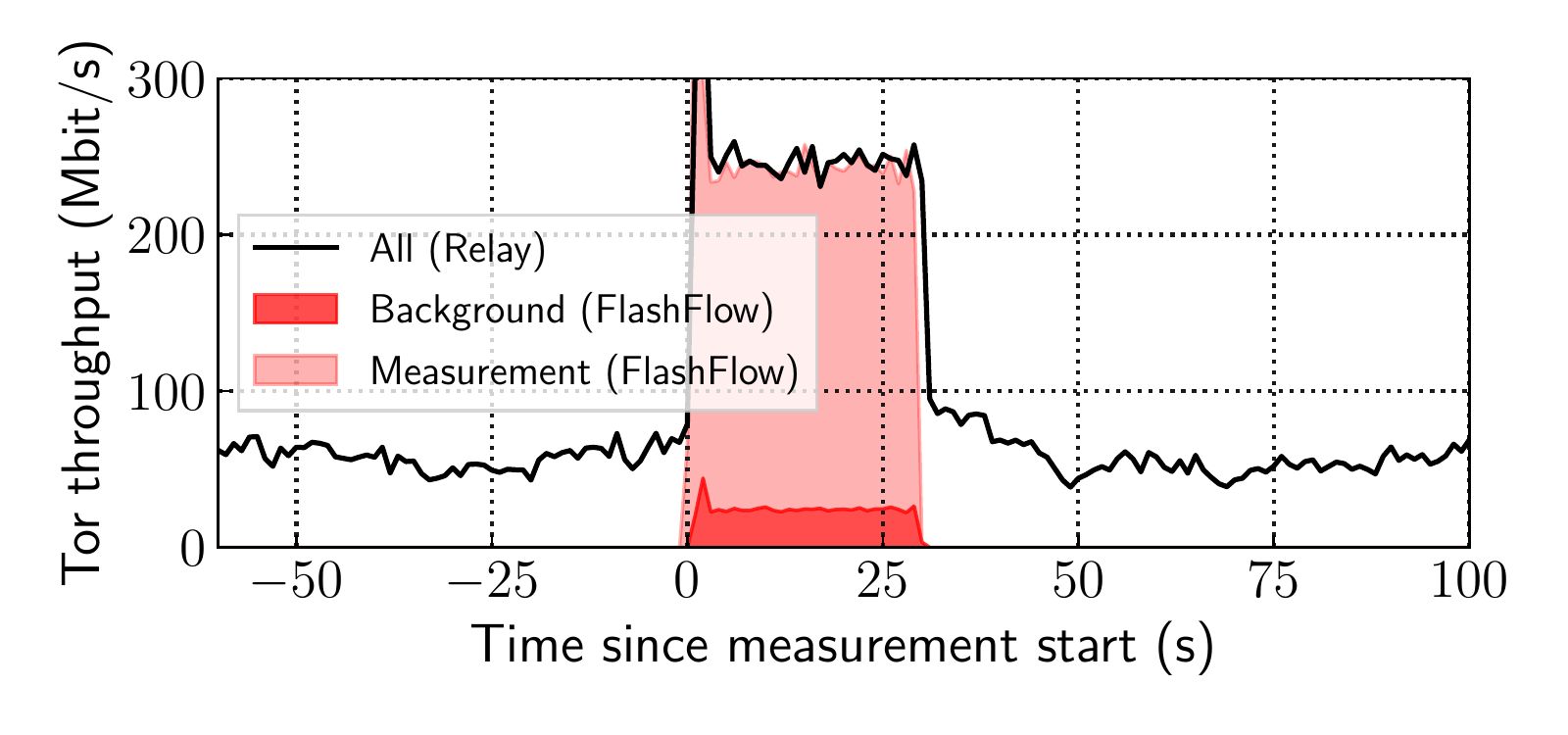}
	\caption{Tor throughput during measurements of a relay with client
	background traffic as reported by the \tool{} and by the relay. The
	shaded regions are stacked, and \tool{} reports their median per-second
	sum as its result.}
	\label{fig:bg-traffic-10}
	\vspace{-3mm}
\end{figure}

Figure~\ref{fig:bg-traffic-10} confirms that the sum of the background and
measurement traffic reported by \tool{} is equal to the total traffic
reported by the relay.  Following the measurements, the relay's
throughput immediately returns to the level it was before,
demonstrating that \tool{} has no lingering effect on background traffic
levels. We also observe in Figure~\ref{fig:bg-traffic-10} that
background traffic is limited to 25 Mbit/s as is expected. Note that
the spike at the beginning of the measurement is due to the Tor relay
allowing a one second burst before limiting its own throughput to 250
Mbit/s.

We observe that since our relay's background traffic level was 50
Mbit/s with a capacity limited to 250 Mbit/s, we would need to
configure $r < 0.2$ in order to cause the relay to withhold
background traffic during the \tool measurement. Because relatively
high capacity relays with low Tor throughput will not need to limit
background traffic, we believe that $r=0.25$ provides a reasonable
trade off for limiting a malicious relay to only inflate their
measurement by $1/(1-r) = 1.33$ (see \S~\ref{sec:security}).

\Section{Simulation Experiments \pagebudget{0.75}} \label{sec:network}

\paragraph{Network Measurement Efficiency}
We evaluate the efficiency of \tool{} in measuring the entire Tor network in terms of
its speed. To estimate these values, we simulate measurement of the network by a single
team. We use a greedy scheduler to determine the fastest that we can measure the entire network.
Then we replay the appearance of new relays in the consensus and determine how efficiently they can
be measured as well.

We determine the state of the Tor network over July 2019 using archived Tor
consensuses and descriptors~\cite{tormetrics}. Similar to \S~\ref{sec:torflow},
we estimate the capacity of relay $r$ at time $t$ to be the minimum of the rate
limits set in the relay's descriptor at $t$ and the largest observed bandwidth
for $r$ in the period June--August 2019. Among all relays, the largest capacity
thus determined for July 2019 is 998 Mbit/s.

We estimate how fast \tool{} could measure the entire network for each day in July 2019. For this
estimate, we use the first consensus in the day, and we assume that all of the relays in the network
have been measured before and thus have capacity estimates. We greedily assign relays to each slot
in order, with each assignment choosing the largest relay for which there is available capacity to
measure. We use a measurement team consisting of 3 measurers with 1 Gbit/s capacity each. This team
has capacity that is just larger than the minimum required to accurately measure the largest relay
seen, which due to the excess factor $f = 2.84$ and maximum capacity of $0.998$ Gbps is 2.84 Gbit/s.

The result for the median day is that 5 hours (i.e. 599 30-second slots) are
needed to measure the entire network, with a minimum of 4.9 and a maximum of
5.1. The schedule measures a median of 6,419 relays (min: 6,355, max: 6,528)
with a median total capacity of 608 Gbit/s (min: 592, max: 621). This speed
suggests that the entire network could be measured at least every 24 hours with
significant spare capacity to measure new relays as they join the network.

We next estimate how quickly new relays can be measured. A relay is considered new if it has not
been seen in the last month. We consider each consensus in July 2019 and assume relays in the first
consensus are not new. During this time, there is a median of 3 new relays in a consensus (min: 0,
max: 98). We use as the new-relay capacity estimate the 75th percentile advertised bandwidth from
descriptors in June 2019, which is 51 Mbit/s. The simulation result is that the median time to
measure new relays in a consensus is 30 seconds (min: 0 minutes, max: 13 minutes). These results
show that new relays can be measured within minutes even while \tool{} re-measures the entire
network every 24~hours.

\begin{figure}[t]
	\centering
	\small
	\captionsetup{skip=2pt} %
    \begin{subfigure}[b]{0.5\columnwidth}
        \centering
        \captionsetup{skip=2pt} %
        \includegraphics[width=1.0\textwidth]{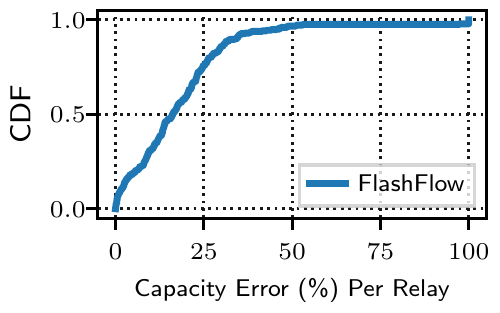}
        \caption{Relay Capacity Error (Eq.~\ref{eq:relaycaperr})}
        \label{fig:shadow:error:rce}
    \end{subfigure}%
    \begin{subfigure}[b]{0.5\columnwidth}
        \centering
        \captionsetup{skip=2pt} %
        \includegraphics[width=1.0\textwidth]{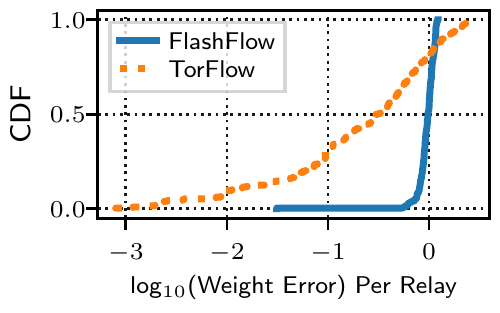}
        \caption{Relay Weight Error (Eq.~\ref{eq:relayweighterr})}
        \label{fig:shadow:error:rwe}
    \end{subfigure}
	\caption{
	Measurement error during concurrent relay measurement in Shadow simulations.
	The corresponding network capacity error\,(Eq.~\ref{eq:networkcaperr}) is 14\% for \tool, while the
	corresponding network weight error\,(Eq.~\ref{eq:networkweighterr}) is 4\% for \tool and 29\% for TorFlow.
	}
	\label{fig:shadow:error}
	\vspace{-3mm}
\end{figure}

\begin{figure*}[t]
	\centering
	\small
	\captionsetup{skip=2pt} %
    \begin{subfigure}[b]{0.61875\textwidth}
        \centering
        \captionsetup{skip=2pt} %
        \includegraphics[width=1.0\textwidth]{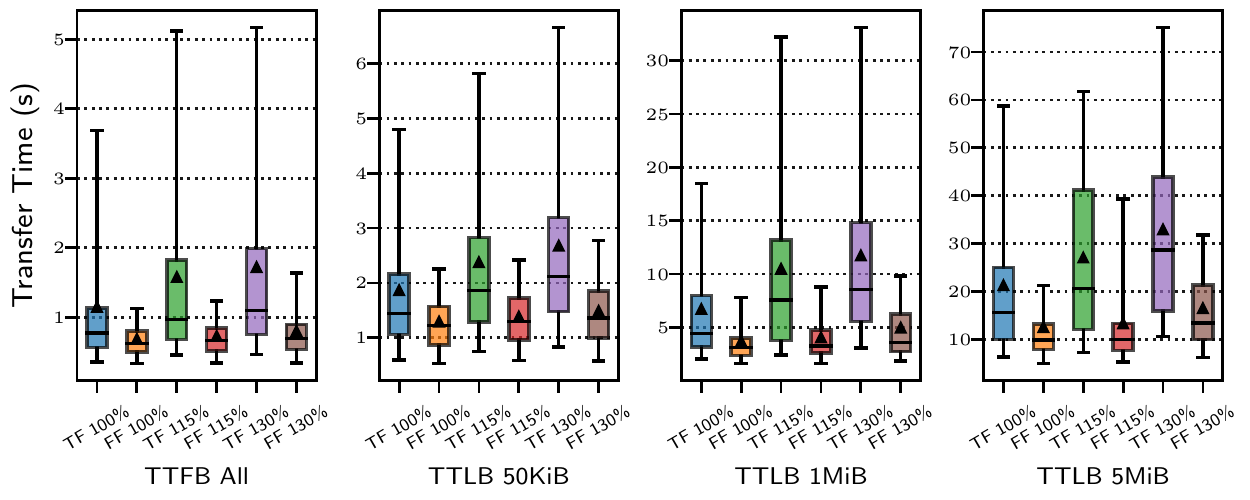}
        \caption{Benchmark Performance}
        \label{fig:shadow:perf:ttb}
    \end{subfigure}%
    \begin{subfigure}[b]{0.185625\textwidth}
        \centering
        \captionsetup{skip=2pt} %
        \includegraphics[width=1.0\textwidth]{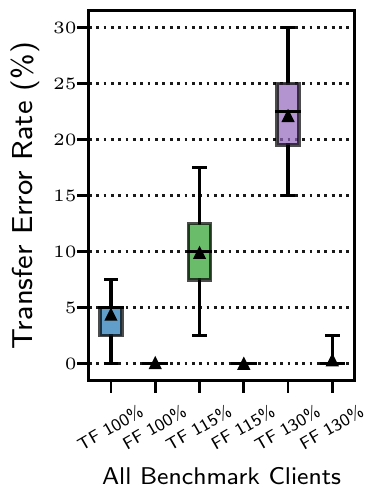}
        \caption{Benchmark Errors}
        \label{fig:shadow:perf:err}
    \end{subfigure}
    \begin{subfigure}[b]{0.185625\textwidth}
        \centering
        \captionsetup{skip=2pt} %
        \includegraphics[width=1.0\textwidth]{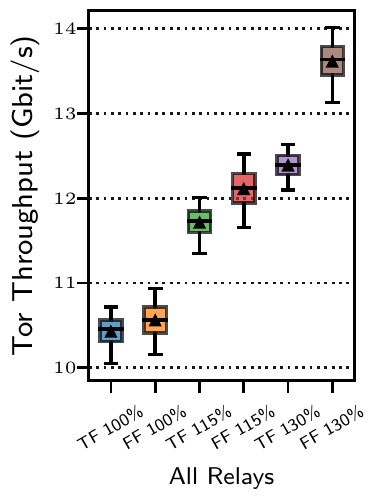}
        \caption{Tor Throughput}
        \label{fig:shadow:perf:tput}
    \end{subfigure}
	\caption{
	Performance results when using TorFlow (TF) and FlashFlow (FF) weights in Shadow simulations with normal (100\%) and extra (115\%, 130\%) traffic load.
	\subref{fig:shadow:perf:ttb} Time to first and last byte of 50\,KiB, 1\,MiB, and 5\,MiB transfers by performance benchmark clients.
	\subref{fig:shadow:perf:err} Fraction of benchmark client transfers that failed (timed out).
	\subref{fig:shadow:perf:tput} Tor network throughput (for every second, sum of relays' Tor throughput).
	In the boxplots, the horizontal line shows the median, the triangle shows the mean, the box shows the interquartile range, and the lower and upper whiskers extend to the 5th and the 95th percentile, respectively.
	}
	\label{fig:shadow:perf}
\end{figure*}

\paragraph{Network Measurement Accuracy and Performance}
We evaluate \tool in a full Tor network deployment using
Shadow~\cite{shadow}, a discrete-event network simulator and a
standard tool for conducting Tor performance
experiments~\cite{torexptools}. We configure a private Tor test
network in Shadow that is 5\% of the size of the public network and
contains: 3 DirAuths; 328 relays; 397 TGen clients that use Tor Markov
models to generate the traffic flows of 40k Tor
users~\cite{tmodel-ccs2018}; and 40 TGen clients that mirror Tor's
performance benchmarking process by repeatedly downloading 50\,KiB,
1\,MiB, and 5\,MiB files (timeouts are set to 15, 60, and 120 seconds, respectively).
The relays were sampled from Tor's consensus
files from January 2019~\cite{tormetrics} and placed in the closest
city in Shadow's Internet map according to IP address and following
best practices~\cite{methodically-modeling-tor}. Each relay is
configured with a capacity equal to the maximum observed bandwidth of
the corresponding relay in the public Tor network during January 2019.

To measure accuracy, we first run a base \tool simulation (using our
implementation and configuration from \S~\ref{sec:deployment}) in
which \tool uses 3 measurers with capacities of 1\,Gbit/s each to measure
the Tor network and produce a bandwidth file containing a capacity
estimate and weight for each relay. We repeat the simulation with
TorFlow, which produces a bandwidth file with weights only (see
\S\ref{sec:background}). We use the capacity estimates, weights, and
the ground truth throughput of each relay to compute the relay and
network measurement errors as described in
Equations~\ref{eq:relaycaperr},\,\ref{eq:networkcaperr},\,\ref{eq:relayweighterr},\,and\,\ref{eq:networkweighterr}.

Figure~\ref{fig:shadow:error} shows the relay capacity and weight error as CDFs over all relays.
Although Figure~\ref{fig:shadow:error:rce} shows that both the median
and inter-quartile range of capacity error \textit{across relays} is
16\%, the corresponding network capacity error (weighted by the
magnitude of the absolute error) is only 14\% \textit{in total}.
Figure~\ref{fig:shadow:error:rwe} compares the relay weight error for
\tool and TorFlow, where $x=0$ represents ideal relay weighting and
each unit on the x-axis represents a 10$\times$ increase in error. We
observe that more than 80\% of relays are underweighted by TorFlow
compared to their ground truth capacity, following our conclusions that were drawn
from Figure~\ref{fig:relayweighterr} in \S\ref{sec:torflow}. \tool
shows considerable improvement in relay weighting, with a total
of only 4\% network weight error (Equation~\ref{eq:networkweighterr})
compared to 29\% for TorFlow.

To measure performance, we use the bandwidth files produced by \tool
and TorFlow in the above simulations to run 3 new simulations for each
system; one simulation is configured with normal (100\%) traffic load,
one with 15\% extra (115\%) traffic load, and one with 30\% extra (130\%) traffic load.
In all simulations, Tor is
configured to form a consensus with the previously measured relay weights,
and therefore client load is balanced according to these weights.

Figure~\ref{fig:shadow:perf} shows considerable improvement in
performance when using the \tool weights compared to TorFlow across
\textit{all metrics and benchmarks}. Figure~\ref{fig:shadow:perf:ttb} shows that the
\tool benchmark clients \textit{outperform} the TorFlow benchmark clients across all transfer sizes:
in the 100\%-loaded simulations, the median of
50\,KiB, 1\,MiB, and 5\,MiB transfer times decreases by 15\%, 29\%,
and 37\%, respectively. \tool also yields more \textit{consistent} client
performance: in the 100\%-loaded simulations,
the standard deviation of 50\,KiB, 1\,MiB, and 5\,MiB transfer
times decreases by 55\%, 61\%, and 41\%, respectively.
We also observe that \tool better supports \textit{network growth}
because the performance improvements increase as the network becomes
more loaded. For example, relative to TorFlow, the median 1\,MiB
transfer time in \tool decreases by an additional 28\% and 29\%
when the network is 15\% and 30\% more loaded, respectively.
Surprisingly, performance in the 130\%-loaded \tool simulation was
still better than performance in the 100\%-loaded TorFlow simulation,
across all transfer sizes.
Figure~\ref{fig:shadow:perf:err} shows that the median rate of
transfer timeouts decreases by 100\% in all \tool simulations,
compared to median transfer failure rates of 5\%, 10\%, and 23\% for
TorFlow in the 100\%-, 115\%-, and 130\%-loaded simulations,
respectively.
Finally, Figure~\ref{fig:shadow:perf:tput} shows that the \tool
weights result in a more \textit{balanced network} that is more
capable of handling additional traffic load. Increasing client-traffic
load by 15\% and 30\% resulted in a 15\% and 29\% increase in the
median Tor throughput (summed over all relays) in \tool as expected,
but only a 12\% and 18\% increase in TorFlow, respectively.
Overall, our simulations demonstrate that \tool is significantly more
capable of balancing load in Tor than is TorFlow.

\Section{Related Work \pagebudget{0.5}}

\paragraph{Load Balancing in Tor}
Several systems for load balancing in Tor have been proposed. Load-balancing
systems produce the relay weights that clients use to select paths, and in
some cases the relay capacities can also be determined. A comparison of
these systems appears in Table~\ref{table:comparison}. It shows the added server
bandwidth required, the demonstrated success factor of a weight-inflation
attack, if the system provides capacity values in addition to weights for
load-balancing, and how long it takes to produce weights for the entire network.
We observe that for some increase in required server bandwidth, \tool{} provides
increased security and speed, and it can be used for capacity estimates as well
as load balancing.

The Tor network currently uses TorFlow~\cite{perry2009torflow} to estimate
relays' capacities and assign weights accordingly. We discuss TorFlow in
\S~\ref{sec:bwbackground} and its limited accuracy in \S~\ref{sec:torflow}.
TorFlow is vulnerable to attacks~\cite{low-resource-attack, thill-thesis,
peerflow-pets2017}, the most straightforward of which is that a malicious relay
can falsely report very high bandwidth information in its
descriptor~\cite{low-resource-attack}, increasing its final weight regardless of
its performance measurements. Such attacks have been demonstrated to increase
the weight of a Tor relay by $89\times$~\cite{thill-thesis} to
$177\times$~\cite{peerflow-pets2017}. Data from TorFlow's
BWAuths~\cite{bwauth-status,trac17482} indicate that a single 1 Gbps scanner
takes at least 2 days to measure the entire network.

SmarTor~\cite{andre2018smartor} decentralizes the operation of the BWAuths using
a blockchain and trusted execution environments. Similar to TorFlow, it measures
a relay's capacity by downloading a file through the relay in a measurement
circuit. It thus remains vulnerable to bandwidth-inflation attacks demonstrated
against TorFlow.
We do not include SmarTor in Table~\ref{table:comparison}
because its contributions over TorFlow are not to the measurement technique
itself. Consequently, its measurement attributes can be assumed to be similar to
that of TorFlow.

EigenSpeed~\cite{eigenspeed} uses a peer-measurement approach in which every
relay records the average per-stream throughput with every other relay and
reports this vector to the Tor DirAuths. The DirAuths combine the vectors into a
matrix and iteratively compute the eigenvector of that matrix as the relay
weights. For security, this computation must be initialized with the weights
from a certain number of trusted relays. During and after the eigenvector
computation, relays can be marked as malicious due to atypical changes in or
unusual final values of their weights, and these marked relays are effectively
removed from the network. EigenSpeed observations are per-flow throughputs
rather than total relay capacity.

EigenSpeed is vulnerable to several attacks~\cite{peerflow-pets2017}. First,
unevaluated relays receive weights of $1/n$, given $n$ total relays, enabling a
Sybil attack to yield disproportionate weights for the malicious relays. Second,
an \emph{increase framing attack} allows an adversary controlling just 2\% of
the network to frame up to 20\% of the honest relays as malicious and have them
removed from the network. Finally, an \emph{targeted liar attack} allows a set
of malicious relays to inflate their total weight to 7.4--28.1 times the weight
they deserve, depending on the number of trusted relays.

\begin{table}[]
\begin{threeparttable}
\captionsetup{skip=0pt} %
\caption{Comparison of Tor load-balancing systems} \label{table:comparison}
\centering
\footnotesize
\begin{tabular}{llllll}
\toprule
& \begin{tabular}[c]{@{}l@{}}\textbf{Server}\\ \textbf{BW}\end{tabular} & \begin{tabular}[c]{@{}l@{}}\textbf{Attack}\\ \textbf{Advantage}\end{tabular} & \begin{tabular}[c]{@{}l@{}}\textbf{Capacity}\\ \textbf{Values}$^{\star}$\end{tabular} & \textbf{Speed} \\
\midrule
\textbf{TorFlow}$^{\dagger}$ & 1 Gbit/s & $177\times$ & \multicolumn{1}{c}{\LEFTcircle} & 2 days \\
\textbf{EigenSpeed} & 0$^{\ddagger}$ & $21.5\times^{\diamond}$ & \multicolumn{1}{c}{\Circle} & 1 day \\
\textbf{PeerFlow} & 0$^{\ddagger}$ & $10\times^{\diamond}$ & \multicolumn{1}{c}{\LEFTcircle} & 14 days$^{+}$ \\
\textbf{FlashFlow} & 3 Gbit/s & $1.33\times$ & \multicolumn{1}{c}{\CIRCLE} & 5 hours\\
\bottomrule
\end{tabular}
\begin{tablenotes}
\footnotesize
\item [$\star$] Values provided (\CIRCLE), can be inferred (\LEFTcircle), or unavailable (\Circle).
\item[$\dagger$] SmarTor can be assumed to have attributes similar to TorFlow.
\item [$\ddagger$] Relays measure each other using existing client traffic.
\item [$\diamond$] With 20\% trusted relays (by number or weight).
\item [$+$] Time to measure largest 96.8\% of relays.
\end{tablenotes}
\end{threeparttable}
\vspace{-3mm}
\end{table}

In PeerFlow~\cite{peerflow-pets2017}, relays periodically report to the DirAuths
the total number of bytes they exchange with each other. The DirAuths then
securely aggregate the traffic data to produce relay weights. In the process of
determining weights, PeerFlow produces lower bounds on relay capacities that
can be used as capacity estimates. PeerFlow requires
a fraction $\tau$ of relay weight that is trusted, and the adversary can obtain
weights for his relays inflated by a factor of $2/\tau$. If $\tau=0$, then a
sufficiently large adversary (i.e. relative weight above 4\%) can eventually get
an arbitrarily large relative weight. PeerFlow also limits how quickly a
malicious relay's weight can increase from one measurement period to the next.
Based on the suggested parameters, a malicious relay can inflate its claimed
capacity by a factor of 4.5 (see Theorem 1 of~\cite{peerflow-pets2017}).

In comparison to these systems,
\tool{} has
much better protection against weight inflation both in the short term and long
term, as it has an inflation factor of 1.33 at all times. It also allows the entire
network to be measured in hours rather than days. \tool{} does require higher
measurement-server bandwidth than previous systems, but the requirement is still
not high (3 Gbit/s), especially compared to the capacity of the Tor network
itself (> 400 Gbit/s).

We note some additional systems superseded by later work or that do not directly
produce load-balancing weights. Snader and Borisov~\cite{snader-ieee2011}
propose a simple form of EigenSpeed's peer measurement that takes the median of
pairwise speed observations. It uses an unweighted median and is thus vulnerable
to a Sybil attack. TightRope~\cite{tightrope-wpes18}, assumes that capacity
weights already exist for the relays and then considers how to choose paths to
optimally balance load. Using differential privacy, the current load on all
relays is shared with a server that computes a distribution for clients to use
when building new circuits. Wang et al.~\cite{wang2012} propose Tor clients use
lightweight active measurements that use latency as an indicator for congestion,
detect congested relays, and automatically avoid using them.

\paragraph{Other Related Work}
Speed tests such as Ookla~\cite{ookla-speedtest} are
primarily intended for home users to test the throughput of their
devices, wireless router, or their ISP's connection.
iPerf~\cite{iperf} can achieve high throughput at the transport layer over both
UDP and TCP. Prasad et al.~\cite{prasad2003bandwidth} describe
bandwidth-estimation techniques, focusing on efficient techniques such as packet
pairs and trains. Feamster and Livingood discuss the challenges of Internet
throughput measurement even when allowing the measurement to fully utilize
bandwidth~\cite{feamster-arxiv2019}.

\Section{Conclusion \pagebudget{0.25}}

Tor's load-balancing system utilizes self-reported capacity estimates from Tor
relays, a process that is vulnerable to malicious reporting. We furthermore show
through an analysis of Tor metrics data that these estimates are significantly
inaccurate and result in suboptimal load balancing. We then present \tool, a
system to actively measure Tor relays with limited effect on normal client traffic. We implement \tool, conduct extensive experiments, and show it
accurately, securely, and quickly measures the capacities of Tor
network relays. Moreover, we show through simulation that these capacities improve the load-balancing of Tor.

Our results show that \tool could be used today to improve Tor's performance and
resource estimates. Furthermore, \tool could be used as a secure basis for
incorporating additional dynamic performance measurements. Such measurements,
such as per-relay network and CPU utilization, could provide information about
\emph{available} (rather than total) capacity that may further improve Tor's
load balancing. The \tool measurements would be used as a starting weight, and
then the weights would only be reduced, depending on the dynamic measurements.
\tool would thus securely limit the weight of any relay while allowing for
improved performance via adjustments based on insecure dynamic measurements,
such as self-measurements.

\section*{Acknowledgments}
\looseness-1
This work was supported by the Office of Naval Research.

{\small
\bibliographystyle{abbrvnat}
\bibliography{paper}
}

\appendix
\Section{Relay Capacity and Weight Variation}\label{sec:torflowvariation}

We extend the TorFlow analysis from \S~\ref{sec:torflow} to better
understand how relay capacity estimates and relay weights vary across
relays and over time.

\begin{figure*}[t]
        \begin{subfigure}[b]{0.5\textwidth}
                \centering
                \captionsetup{skip=0pt} %
                \includegraphics[width=.8\textwidth]{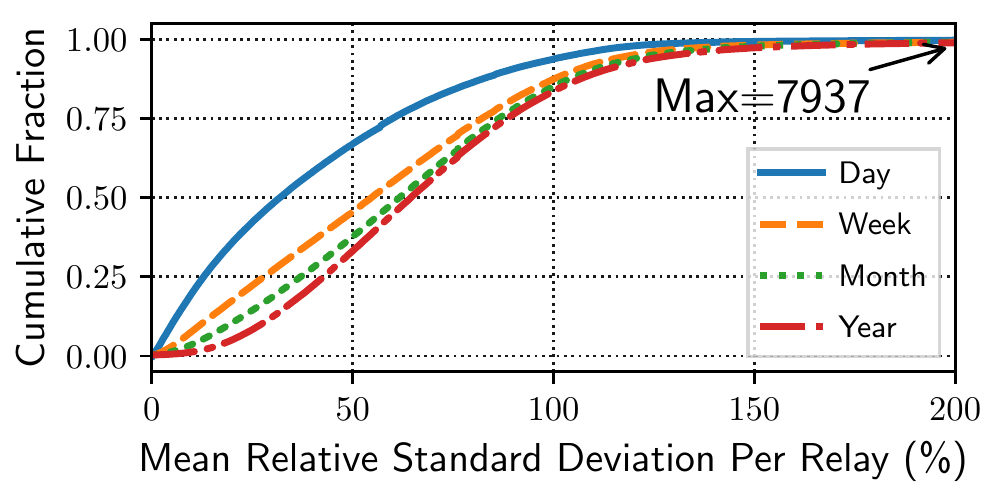}
                \caption{Relay Capacity Varation CDF}
                \label{fig:relaycapvar}
        \end{subfigure}%
        \begin{subfigure}[b]{0.5\textwidth}
                \centering
                \captionsetup{skip=0pt} %
                \includegraphics[width=.8\textwidth]{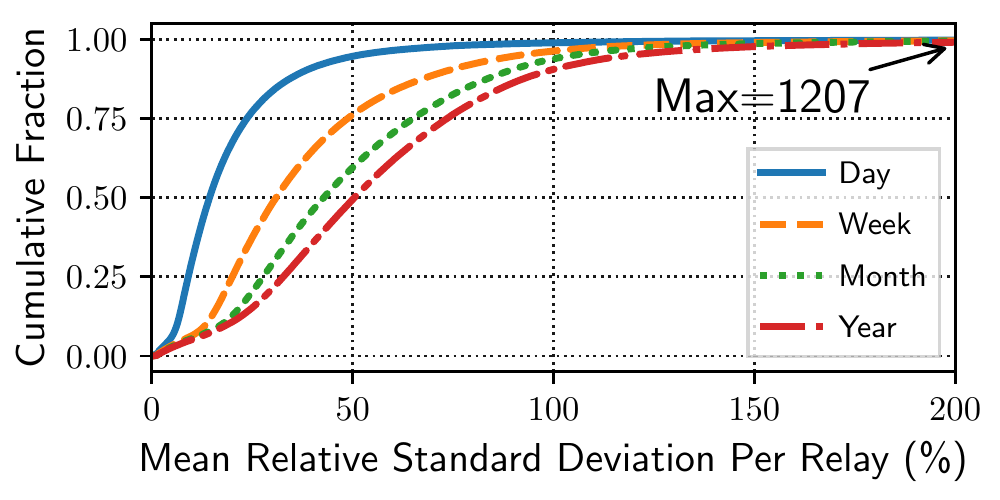}
                \caption{Relay Weight Varation CDF}
                \label{fig:relayweightvar}
        \end{subfigure}
	\vspace{-3mm}
	\caption{
	Variation relay capacities and weights over time, computed using 11 years of archived Tor metrics data~\cite{tormetrics}.
	}
	\label{fig:variation}
	\vspace{-3mm}
\end{figure*}

\paragraph{Relay Capacity Variation}
A relay with a perfect capacity estimation algorithm would consistently report
the same advertised bandwidth. Variation in advertised bandwidths thus indicates
inaccurate capacity estimation, which further leads to poor load balancing.
Moreover, the resulting variable load leads to unpredictable and frustrating
client performance. We use the \textit{relative standard deviation} (RSD) as a
measure of the variability in relays' advertised bandwidths over time. We
compute the RSD over a sequence of values $V$ as
\begin{equation} \label{eq:relstddev}
\textrm{RSD}(V) = \textrm{stdev}(V) / \textrm{mean}(V),
\end{equation}
where \textrm{stdev}() and \textrm{mean}() compute the standard deviation and
mean, respectively.

Recall from \S~\ref{sec:torflow}
that $A(r, t)$ is the advertised bandwidth of relay $r$ at time $t$, and that
$A(r,t,p)$ is the multiset of advertised bandwidths published during the period
of length $p$ preceding time $t$. Varying the length of time $p$ allows us to
examine variability over different timescales. We summarize this variability for
relay $r$ by computing the mean of $\textrm{RSD}(A(r, t, p))$ over the times $t$
that are on the hours between 2009-08-01 and 2019-07-31 (we start on 2009-08-01
to provide up to a year of data for periods ending then). The distribution of
these means over all relays $r$ appears in Figure~\ref{fig:relaycapvar} for
periods $p$ of 1~day, 1~week, 1~month, and 1~year.

Figure~\ref{fig:relaycapvar} shows that the advertised bandwidths reported by
relays exhibit significant variation. For the median relay $r$, $\textrm{RSD}(r,
p)$ is 32\% when computed for a period length $p$ of a day, and it is 55\%,
62\%, and 65\% when computed for a period length $p$ of a week, month, and year,
respectively. Within a given day, there will only be 2–3 descriptors published
because they are published every 18 hours. There is thus a surprising amount of
variation over a day, which is almost certainly in error due to the very short
time interval. For a week, where we also would not expect much change in true
capacity, the RSD is 27\% or greater for 75\% of the relays, and 82\% or greater
for 25\% of relays. The maximum RSD over a year is 7,937\%. Advertised bandwidth
thus frequently varies a non-trivial amount, and most of this variation is
unlikely to be from genuine changes in relay capacity.

\paragraph{Relay Weight Variation}
Recall that the probability that a relay is selected in a circuit is roughly its normalized
consensus weight, that is, its fraction of the total weight assigned to
all relays. Recall also that $W(r, t)$ represents this value for relay $r$ at time $t$, and
that $W(r,t,p)$ is the multiset of these values over the consensuses during the
period of length $p$ preceding time $t$. To measure the variability of
normalized consensus weights, we again use the RSD. We compute
$\textrm{RSD}(W(r, t, p))$, following Equation~\ref{eq:relstddev}, and summarize
the results over time by computing its mean over all times $t$ on the hours
starting from $t$~=~2009-08-01.

\looseness-1
Figure~\ref{fig:relayweightvar} shows the mean RSD per relay for
various periods $p$. We observe similar trends as we observed for
advertised bandwidths: variation increases as we include more
consensus weights in the RSD computation. The RSD for the median relay
is 14\%, 31\%, 43\%, and 50\% when computed using weights from the
preceding day, week, month, and year, respectively. We also find that
the RSD for 25\% of relays is greater than 23\% when using weights
from the preceding day, and greater than 73\% when using weights from
the preceding year. These results indicate that there is also
significant variation in normalized consensus weights. Note that while
changes in relays' advertised bandwidths will affect their normalized
consensus weights, relays joining and leaving the network (i.e.,
churn) may also affect the weights to some extent.

\Section{Capacity of Internet Hosts}
\label{sec:internettput}

We measure the network performance of our Internet hosts (first described in
\S~\ref{sec:deployment}) with particular focus on the
links between them and US-SW.
Over the course of a day we
run 24 bidirectional iPerf~\cite{iperf} TCP and UDP measurements for 60 seconds
between each host and US-SW and record iPerf's per-second send and receive
throughput statistics.
To summarize the results, at each second we take the minimum of the amount of
sent and received data, take the median of these 60 per-second data points, and
list the range of each host's 24 medians in the first two columns
of Table~\ref{tab:iperf-host-summary}.

In all cases the maximum UDP iPerf throughput is higher than the TCP iPerf
throughput, which is expected because UDP doesn't hold itself back during
perceived packetloss and it has fewer headers (thus less overhead).
The range of TCP iPerf throughput for all hosts except US-NW includes 800
Mbit/s.
US-NW's TCP iPerf throughput is
highly varied, and upon inspection of the results, we determine the variability
is only in its receive direction; when only considering the send direction,
these results show US-NW is capable of sending TCP iPerf traffic at 926--934
Mbit/s.

\begin{table}
\centering
\footnotesize
\begin{threeparttable}
\captionsetup{skip=0pt} %
\caption{Throughput estimation of Internet hosts using iPerf}
\label{tab:iperf-host-summary}
\begin{tabular}{ r | c  c  c}
\toprule
	           & \textbf{TCP (Mbit/s)$^\star$} & \textbf{UDP (Mbit/s)$^\star$} & \textbf{UDP (many)$^\dagger$\,\,\,\,\,\,\,} \\ %
\midrule
	\textbf{US-SW}      & -        & -        &  954  \\
	\textbf{US-NW}      & 176--787 & 740--945 &  946 \\
	\textbf{US-E}       & 874--919 & 943--944 &  941 \\
	\textbf{IN}         & 677--819 & 925--955 & 1076 \\
	\textbf{NL}         & 827--880 & 952--956 & 1611 \\
\bottomrule
\end{tabular}
\begin{tablenotes}
\footnotesize
\item[$\star$] Range of 60-second median iPerf throughput measured by US-SW.
\item[$\dagger$] 60-second median iPerf throughput when saturated by all other hosts.
\end{tablenotes}
\end{threeparttable}
\vspace{-5mm}
\end{table}

As these are measurements between a pair of hosts, either host could be a
bottleneck.  Thus we perform a set of experiments where for each host we
instruct all other hosts to perform a UDP iPerf measurement to it at the same
time for 60 seconds. We sum together the per-second results from each host and
present the median of the summed per-second results in the last column of
Table~\ref{tab:iperf-host-summary}. All three of the US hosts are clearly
limited to about 1 Gbit/s. IN and NL achieve higher throughputs, and while
their hosting provider makes no claim about their capacity, they must have
faster than 1 Gbit/s NICs.

\Section{Tor Processing Limits \pagebudget{1}}\label{sec:tors-limit}

In this section, we provide additional details on the experiments we
ran in \S~\ref{sec:processlimits}
to determine Tor's processing limitation and its effect on the
throughout a relay can achieve.

\Subsection{Setup}
To test Tor, we set up a small test Tor network in a lab environment. On the
\emph{target machine} we run the core Tor network, which including one relay
that we choose to be the target of our tests. On the \emph{client machine} we
run 100 Tor clients, 100 Tor exit relays, a varied number of curl clients, and
an nginx webserver.  Both machines have 2 Xeon E5-2697V3 CPUs with a maximum
frequency of 3.60 GHz, a total of 56 threads, and 256 GiB of RAM. A 10 Gbit/s
fiber cable connects them directly.  The RTT between the machines is 0.13 ms.

Because Tor's KIST scheduler is incapable of fully utilizing a high capacity
link when it has a a small number of active sockets~\cite{trac29427}, we
investigate the relationship between a relay's number of active sockets and its
observed throughput. For 100 \emph{socket} experiments, we instruct $n\in [1, 100]$
Tor clients to each build a two hop circuit beginning at the
target relay and with the $i$th client's circuit ending at the $i$th Tor exit.
We proxy curl processes that download a large
file from the webserver over these Tor clients' circuits, and use three curls
per circuit (Tor's circuit level flow control prevents too much in flight traffic
on a circuit at once, but by running at least two
application streams, one will max out the circuit's flow
control limit).  Note each client opens its own socket to the target relay,
thus the target has $n$ busy open sockets to clients.

Regardless of the number of sockets, more circuits should prevent Tor's
flow control from inducing a
limit, so we also investigate increasing the number of active circuits
on a single socket.  For 100 \emph{circuit}
experiments we instruct a single Tor client to build $n\in [1, 100]$
circuits beginning at the target relay and with the $i$th circuit ending at the
$i$th Tor exit for $i\in [1, n]$. We again proxy three curls over each
circuit to fully utilize each circuit's flow control. Note there is one
client, thus the target always has a single busy open socket to a client.

\Subsection{Results}
\begin{figure}
	\centering\includegraphics[width=0.45\textwidth]{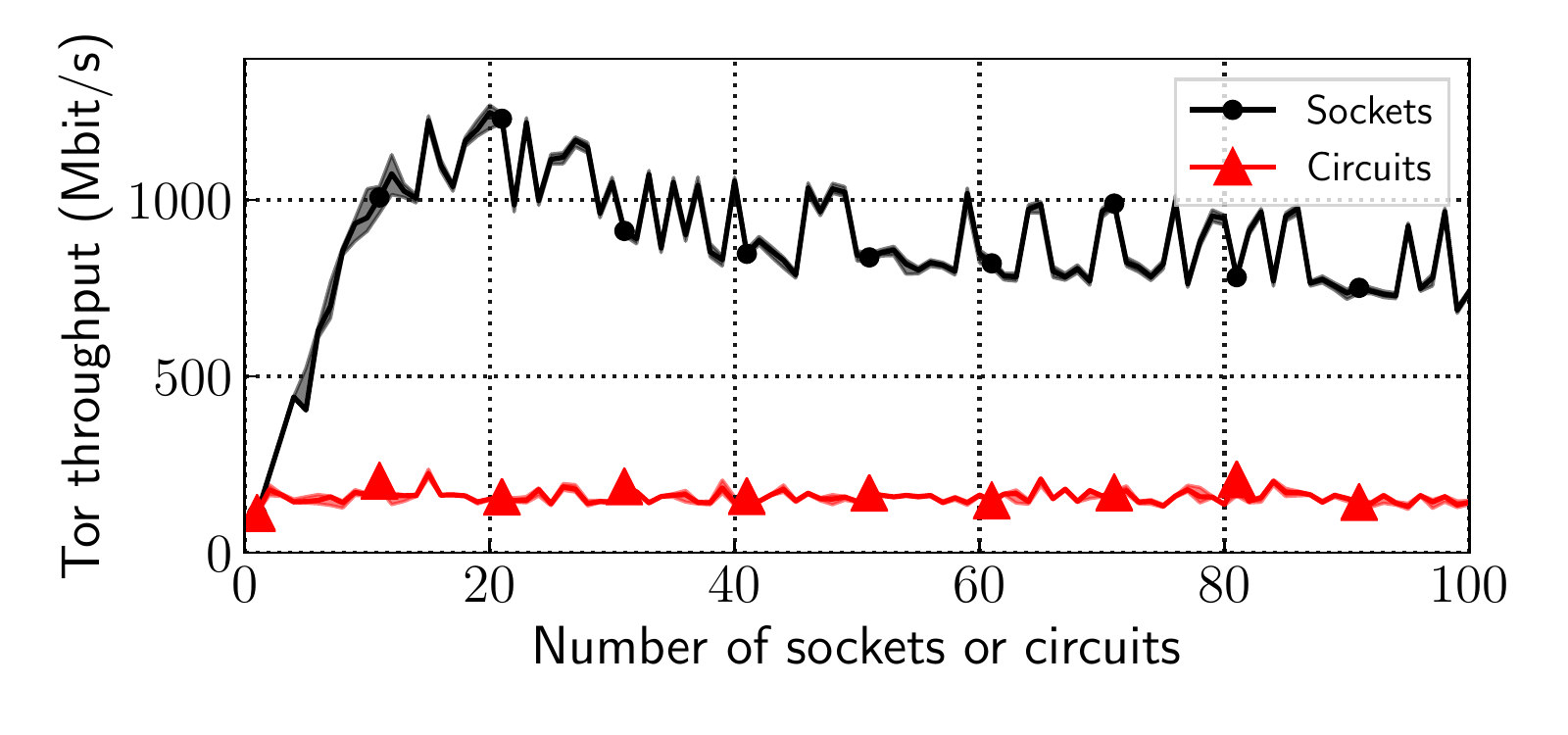}
	\vspace{-3mm}
	\caption{Tor throughput at the target relay, varying the number of
	sockets or circuits. The solid lines are the median per-second
	throughputs from each 120-second experiment, and the small shaded
	regions surrounding the lines are the interquartile ranges.}
	\label{fig:tors-limit}
	\vspace{-5mm}
\end{figure}
Figure~\ref{fig:tors-limit} shows our results from these two sets of tests.
For each number of sockets or circuits, we collect per-second Tor
throughput at the target using Tor's BW events for 120 seconds.  The
median per-second throughputs are plotted together as a solid line. The
interquartile ranges nearly imperceptible and plotted as shaded regions above
and below each experiment's solid line.
While we expected increasing the number of circuits in the \emph{circuits}
experiments would increase the throughput at the target, we suspect
KIST's single socket throughput limitation prevents
additional circuits from increasing throughput.

With 20 sockets in the \emph{sockets} experiments we see the maximum median
per-second throughput of 1,248 Mbit/s at the target relay.  Tor first consumes
100\% usage of a CPU core with 13 sockets, and continues to do so at all higher
numbers of sockets. Due to Tor's primarily single-threaded nature, we take this
as the ground truth of a relay on this hardware.  In the real Tor network, the
relay with the highest observed throughput in July 2019
claimed to have forwarded 998 Mbit/s~\cite{tormetrics}.
This may not be a good estimate of the relay's actual capacity
(e.g. because it never receives enough client
load to reach its capacity),
but it establishes
a lower bound. With the likely differences in hardware and the
limitations of relying on real Tor client throughput demands in mind,
1,248 Mbit/s is a likely approximation of the maximum Tor capacity
today. CPUs with faster clock speeds allow for higher Tor
capacities, and Tor adding support for multithreaded scheduling
would drastically change relays' capacities.

\begin{full}
\Section{TCP Socket Tuning} \label{sec:kernel-tuning}

We investigate tuning the Linux kernel's TCP socket parameters to better
support high capacity and high RTT links. With a tuned kernel, theoretically
\tool{} can use fewer measurement sockets across its measurers (its $s$
parameter from Appendix~\ref{sec:num-socks}), and fewer
sockets would result in less bookkeeping overhead. We first determine the
single-socket throughput improvements of a tuned kernel in the lab. Then on the
Internet we investigate the effect a tuned kernel has as compared to a default
kernel as the number of sockets increases.

\Subsection{Single-socket throughput}\label{sec:tcp-tuning}

Notice Figure~\ref{fig:tors-limit} shows in the \emph{sockets} experiments that
opening additional sockets after the maximum Tor throughput is reached lowers
the achievable throughput. Figure~\ref{fig:num-socks} confirms the
same behavior in a different experiment on the Internet.  We expect a similar
trend affects software in general as the cost of managing many sockets
decreases the time available to forward traffic over them and as an increasing
number of TCP sockets increasingly interfere with each other for a share of the
available link capacity. Regardless, \tool{} is partially implemented in Tor
and is subject to this observed behavior. We are thus motivated to maximize the
throughput of a single \tool{} measurement socket to keep the necessary number
of measurement sockets low.

A major limiting factor on a single socket's throughput is the amount of kernel
socket buffer memory it is allowed to consume and whether that adequately
supports the Bandwidth Delay Product (BDP) of the link.  A link's BDP is its
network capacity multiplied by its RTT. Despite the high 10 Gbit/s capacity of
the link in our lab, its very short 0.13 ms RTT keeps its BDP small at 0.155
MiB.  1 Gbit/s links are increasingly common on the Internet, and a 5th
percentile RTT of 27.4 ms~\cite[\S 6.1.1]{tmodel-ccs2018} on such a link has a
BDP of 3.26 MiB, and a median RTT of 118 ms on a 1 Gibt/s link has a BDP of
14.1 MiB.  Higher BDP values for a link mean the hosts on either end must be
willing to buffer more ``in flight'' data to keep the link throughput near
capacity and to wait for acknowledgments.  Linux picks default socket buffer
parameters on boot based on the host's available system memory; on all our
hosts in the lab and on the Internet Linux chooses 4 MiB and 6 MiB for the read
and write buffers' maximum sizes respectively. For the experiments in this
section, we consider this the \emph{default} kernel parameters, and we consider
a second set of \emph{tuned} kernel parameters with a 64 MiB maximum size for
both reading and writing.

We run tests with these two configurations on the same pair of lab hosts as
described in Appendix~\ref{sec:tors-limit}, but add latency between them using netem~\cite{netem} and
vary the amount added to cover the 5th through 95th percentile in the Internet
RTT dataset \cite[\S 6.1.1]{tmodel-ccs2018}.  On the target host we run a Tor relay, and on the client host we
run \tool{} with a single measurement socket and configured to measure the
target for 240 seconds.
Figure~\ref{fig:tcp-tuning} presents the results as CDFs over each
measurement's 240 per-second data points.

As expected, at all RTTs the tuned kernel measurements achieve higher
throughput than their corresponding default kernel measurements. Also notice
how---for both kernel configurations---as RTT (and thus BDP) increases, the
throughput for that kernel decreases as expected. \tool{} impressively achieves
a maximum median throughput of 1,269 Mbit/s, which is consistent with what
we find as Tor's capacity in Appendix~\ref{sec:tors-limit}.

This shows \tool{} can achieve extremely high 1 Gbit/s throughput with a single
socket and a tuned kernel on Internet links with the Internet dataset's median
of 118 ms RTT or less; however, both machines in this setup need to tune their
kernels, and we cannot expect Tor relay operators to do this.  One can
indirectly increase the amount of buffer space available in the kernel by using
multiple sockets instead of just one, which we now explore.

\begin{figure}
	\centering\includegraphics[width=0.45\textwidth]{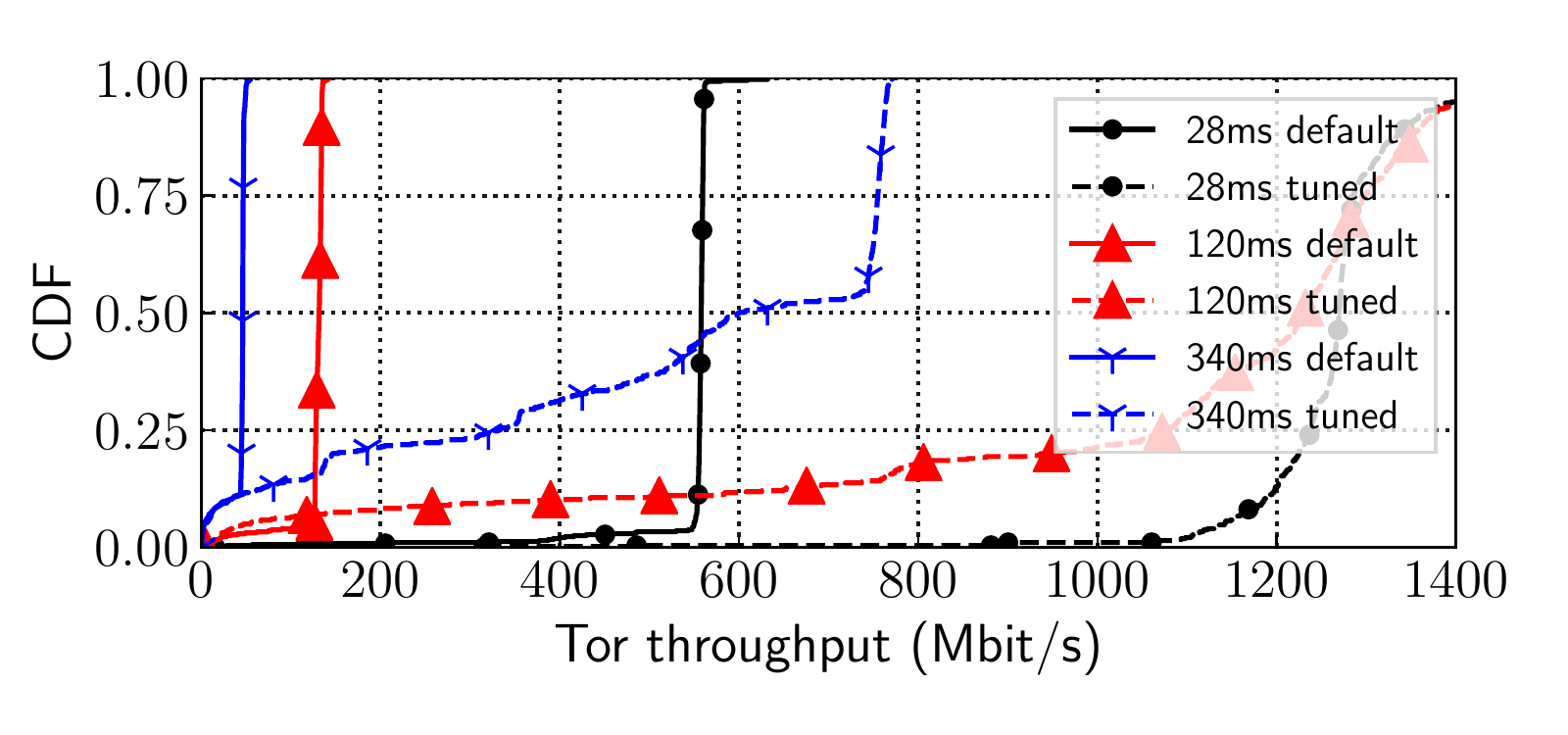}
	\caption{Tor throughput at target relay, presented as CDFs of each
	measurement's 240 per-second throughput data points as measured by
	\tool{} with a single measurer socket, with either the default or tuned
	kernel, and with varied RTT between the target and measurer.}
	\label{fig:tcp-tuning}
\end{figure}

\Subsection{Multi-socket throughput}\label{sec:def-opt-kernel}

\begin{figure}
	\centering\includegraphics[width=0.45\textwidth]{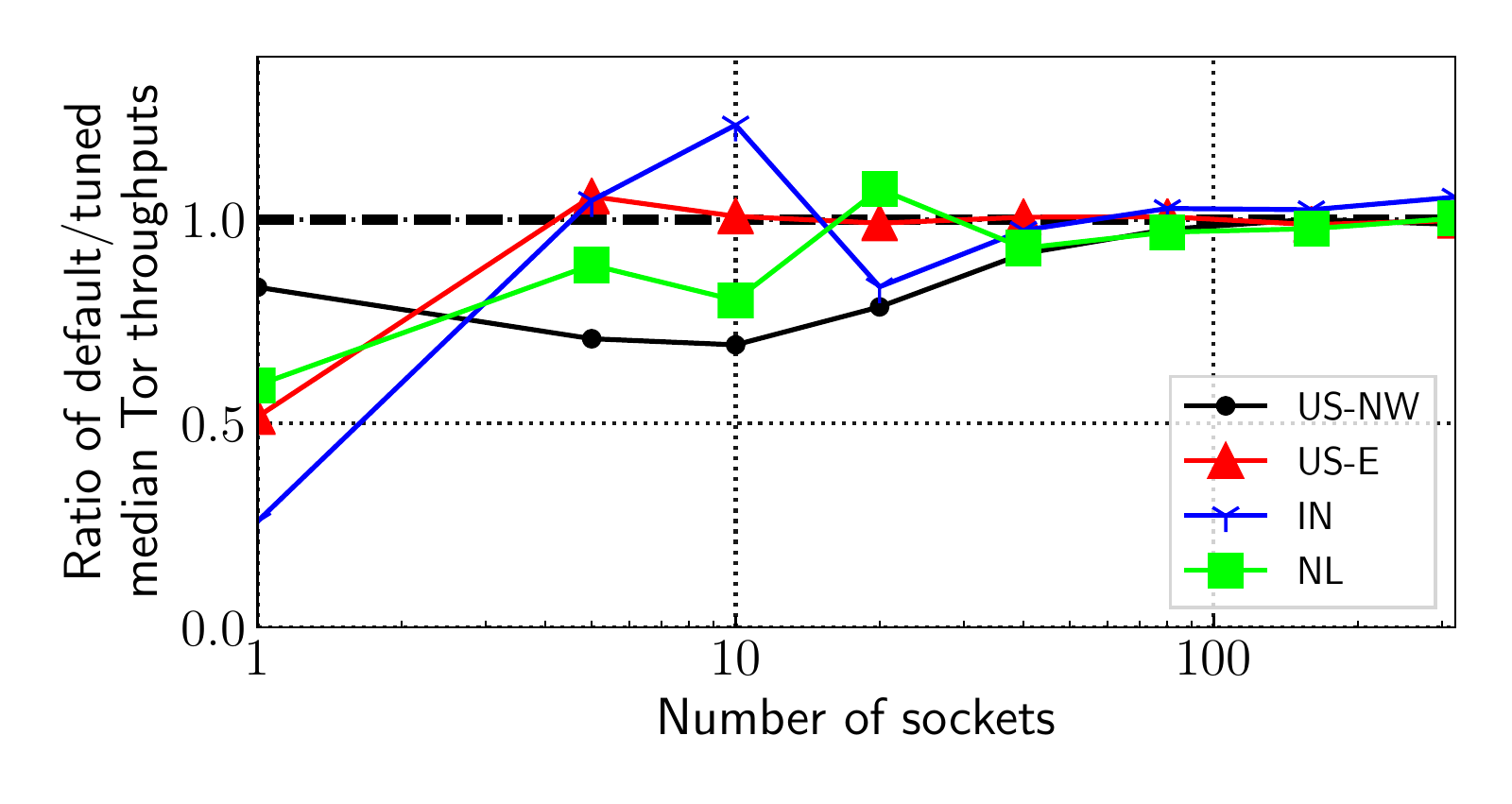}
	\caption{Comparison of default kernel and tuned kernel results,
	presented as ratios. A ratio less than 1 means the tuned kernel
	helped relative to the same experiment with the default kernel.
	Tuning the kernel has less of an effect as sockets are added, thus
	the ratio approaches 1.  The x-axis is log scale.}
	\label{fig:def-opt-kernel-diff}
\end{figure}

Having established a tuned kernel improves single-socket throughput in the lab,
we now move to the Internet to investigate how much tuning the kernel helps as
the number of measurement sockets increases.
At each measurement host individually we use
\tool{} to measure US-SW for 60 seconds, and we consider the case where each uses its
\emph{default} kernel and when each uses its \emph{tuned} kernel. For
presentation we divide each \emph{default} measurement's result by the
corresponding \emph{tuned} result and plot these ratios in
Figure~\ref{fig:def-opt-kernel-diff}.

In all cases, for a small number of sockets a tuned kernel results in a higher
median throughput than the default kernel, as indicated by ratios less than 1.
As the number of sockets used increases, however, tuning the kernel has less of
an benefit, and the plots trend towards 1. This is because the amount of memory
the kernel allocates to buffer traffic for the increased number of sockets
is---in aggregate---able to support the full BDP of the link, nullifying the
benefit of allowing larger buffers per socket in the tuned measurements.

\end{full}
\Section{Deriving Values for \tool Parameters}\label{sec:derive-params}

Before beginning to measure with \tool{} we run a sequence of experiments to
determine appropriate values for its various parameters.
First we determine the number of sockets $s$ the measurement hosts should open
to measure the target relay,
then we explore which multiplier $m$ \tool{} should use when reserving measurer
capacity,
we evaluate various measurement durations $t$ and their accuracies,
and finally we choose error bounds $\epsilon_1$ and $\epsilon_2$.

\subsection{Number of measurement sockets}\label{sec:num-socks}
\begin{figure}
	\centering\includegraphics[width=0.45\textwidth]{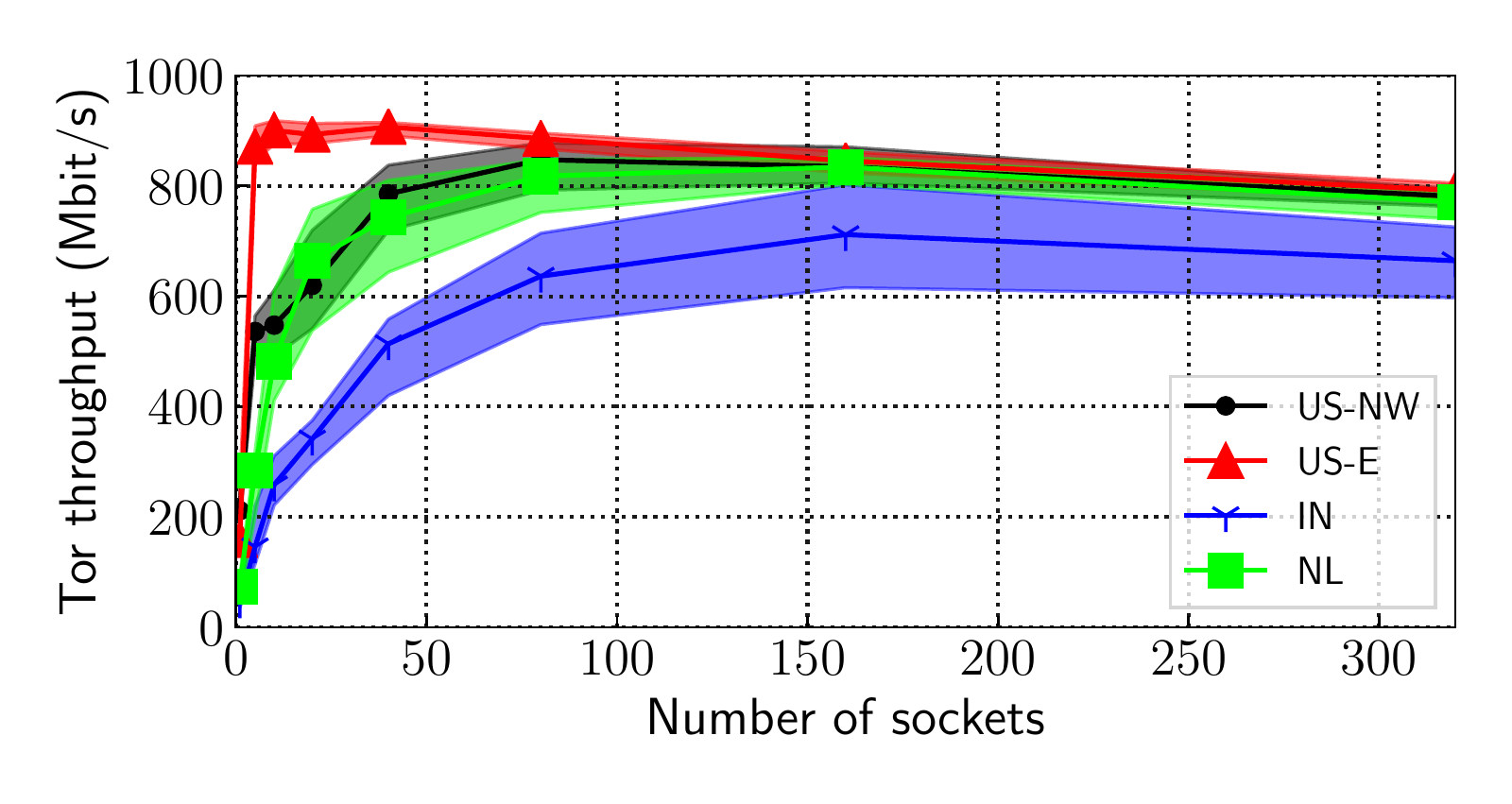}
	\caption{Tor throughput at target relay on US-SW as measured by the
	other machines, varying the number of sockets. Solid lines are median
	per-second throughput, and the shaded regions are the interquartile
	range. Default kernel parameters are used.}
	\label{fig:num-socks}
	\vspace{-5mm}
\end{figure}
We now determine a number of sockets $s$ that, in aggregate, \tool{} measurers
should open to the target.  We use \tool{} to measure US-SW for 60 seconds with each
measurement host pair wise, varying the number of sockets. We do this
until the slowest measurement host stops increasing its throughput.

Figure~\ref{fig:num-socks} shows our results. While each host peaks at a
different number of sockets, IN is the slowest one to peak, and does so at 160
sockets. Thus for all future \tool{} measurements we set the number of sockets
$s$ to use across all measurers to 160.

\looseness-1
We also observe the relative performance of our hosts in
Table~\ref{tab:host-summary}.
We suspect IN produces
the slowest measurements because of its high RTT to US-SW, which generally correlates
with packet loss and therefore lower throughput, as well as its shared virtual
hosting environment in which we do not know and cannot control how many other
virtual hosts share its physical host and compete for its CPU and network
resources.  We suspect the drop in measured throughput after a host's peak is
additional CPU overhead of managing multiple sockets.
US-E is the only non-virtual host, and it performed better than the others.

\subsection{Multiplier}\label{sec:mult}
\begin{figure}
	\centering\includegraphics[width=0.45\textwidth]{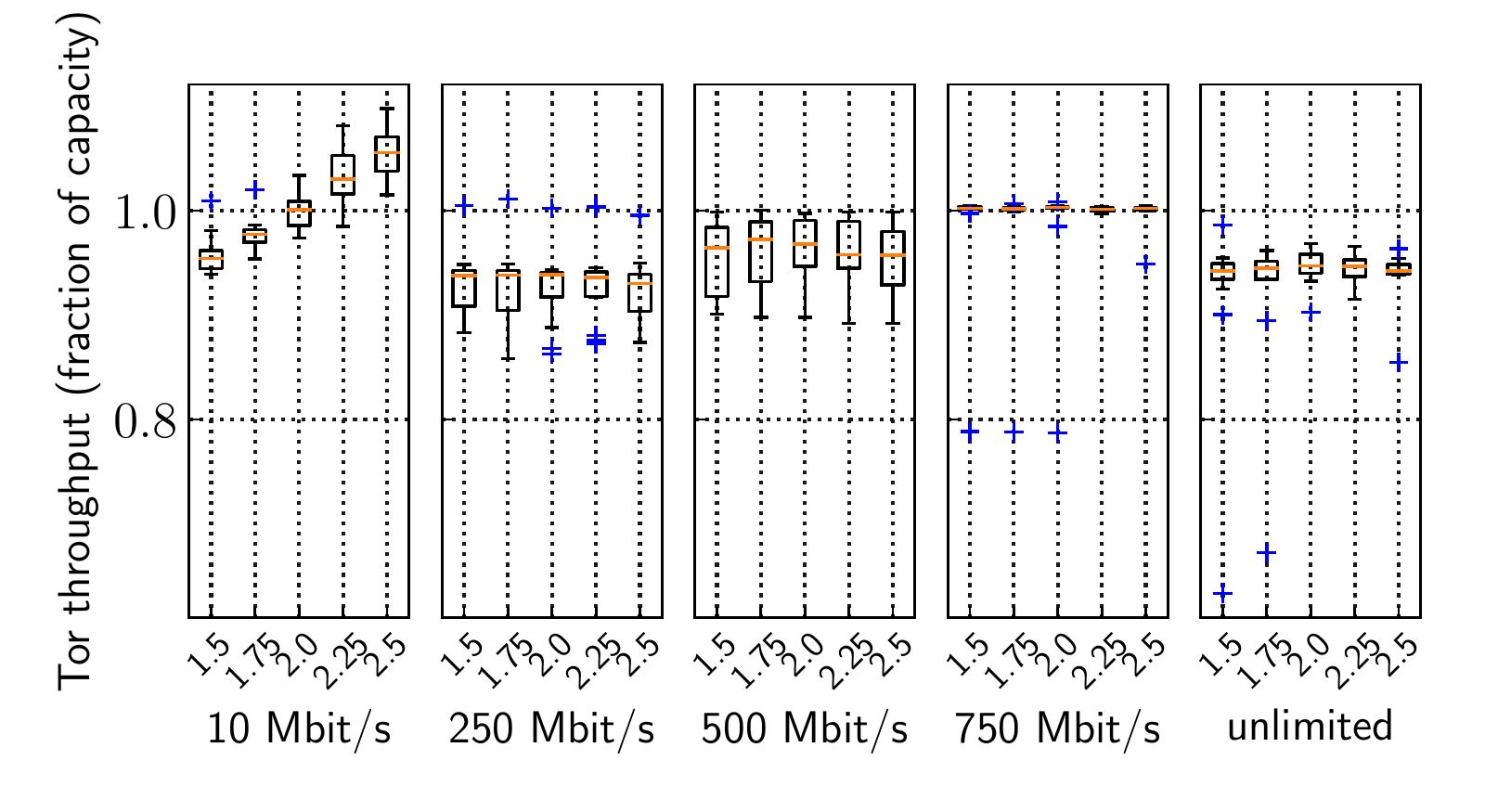}
	\caption{Relative Tor throughput as measured by \tool{} at varied configured
	limits and with a varied multiplier, presented as a fraction of
	ground-truth Tor capacity. Boxplots contain all 60 second medians from
	all subsets with the given multiplier and target capacity.}
	\label{fig:mults}
	\vspace{-5mm}
\end{figure}
As \tool{} runs measurements in parallel, it needs to allocate some amount of
Tor capacity at its measurers for each measurement. Recall from
\S~\ref{subsec:measuring-relay} that \tool{} allocates some factor of measurer
capacity greater than the relay's existing capacity estimate. This factor
depends on a multiplier $m$ for which we now experimentally determine the
smallest value that provides sufficient accuracy.

In these experiments we approximate relays of varied capacities by setting
Tor's \textsf{RelayBandwidthRate} and \textsf{RelayBandwidthBurst} options to
10, 250, 500, and 750 Mbit/s. Many relay operators use these options limit
throughput, and this also simulates relays with such limits at the
network layer.

To establish the ground-truth Tor capacity of these throughput-limited relay
configurations on US-SW, we rerun the two-hop Tor circuit experiment setup
described in Appendix~\ref{sec:tors-limit} and \S~\ref{sec:environment}.
At a Tor throughput limit of
10 Mbit/s, we determine a ground-truth Tor capacity of 9.58 Mbit/s, at
250 Mbit/s ground truth is 239 Mbit/s, at
500 Mbit/s ground truth is 494 Mbit/s, and at
750 Mbit/s ground truth is 741 Mbit/s.
Recall the ground truth of an unlimited relay on
US-SW is 890 Mbit/s.

\looseness-1
Having established ground truth, we now consider multipliers of 1.5, 1.75, 2.0,
2.25, and 2.5.
For each multiplier and at all capacities, we determine all subsets of
measurers with enough measurer capacity to measure the relay, and then we
divide that capacity assignment evenly across the measurers in the subset.
As an example of a subset, to measure a 500 Mbit/s
relay with a multiplier of 1.5 with US-E and IN, we would configure both to
limit their throughput to $\frac{494 \times 1.5}{2}$ Mbit/s.  Limiting
throughput of \tool{} measurers is accomplished using the
\textsf{BandwidthRate} and \textsf{BandwidthBurst} Tor options.

We present the distribution of results at each target relay throughput limit
and each multiplier in Figure~\ref{fig:mults}, normalizing the results as a
fraction of their ground-truth Tor capacities. The lowest multiplier that
avoids outliers below 80\% of ground truth is 2.25. While it has the widest
range of results in 10 Mbit/s measurements, the absolute size of this spread is
only about 0.8 Mbit/s, which is still quite accurate in absolute terms. For
these reasons we choose a multiplier of 2.25 and use it in all future
experiments.

\subsection{Measurement duration}\label{sec:stop-cond}
\begin{figure}
	\centering\includegraphics[width=0.45\textwidth]{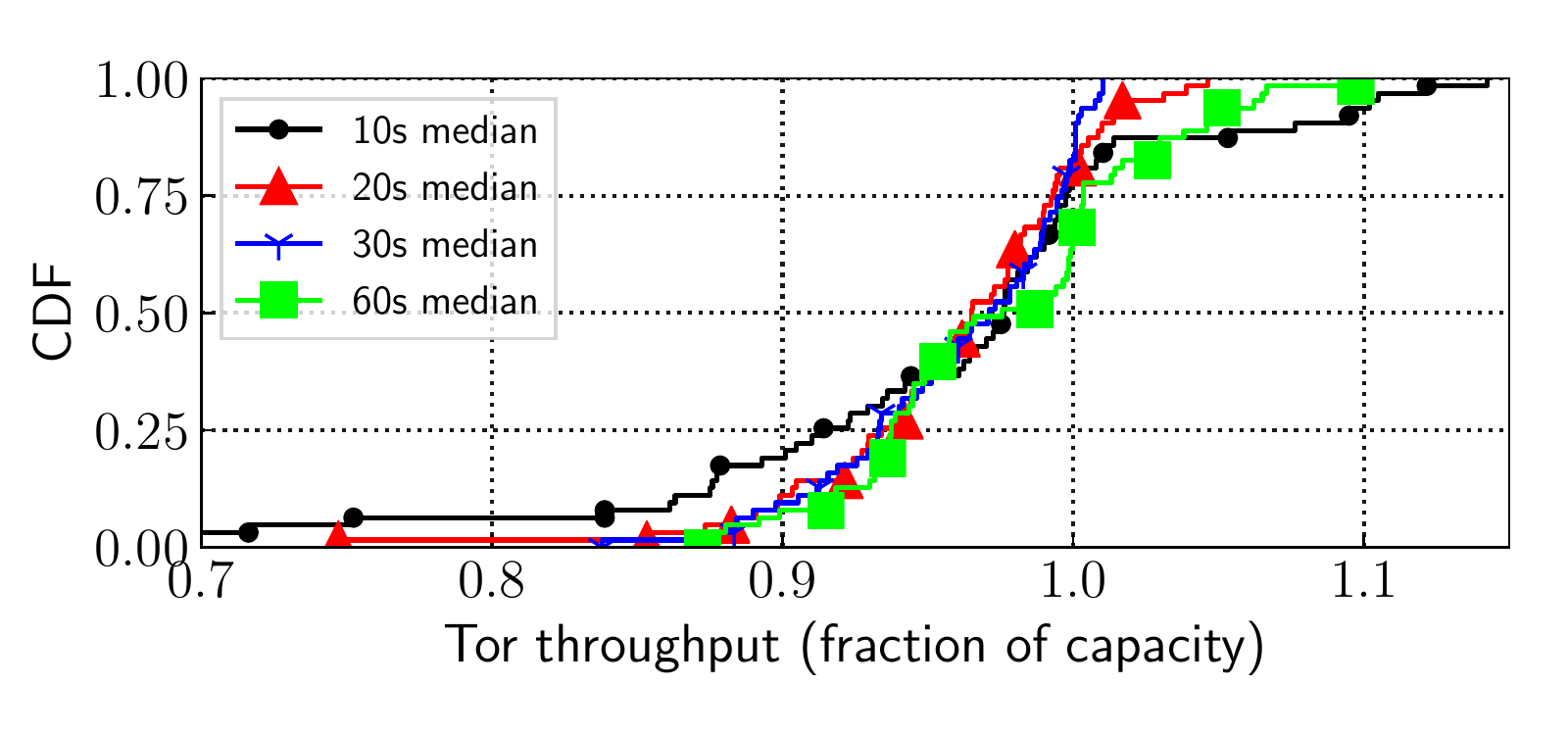}
	\caption{Comparison of accuracy of different measurement lengths.}
	\label{fig:stop-cond}
	\vspace{-5mm}
\end{figure}
Having \tool{} measure for a shorter time would not only allow it to measure
the entire Tor network faster but also prevent it from degrading Tor users'
experience for too long when they have a circuit going through a relay being
measured.
We have been running \tool{} for 60 seconds in the preceeding sections.
We are thus motivated to find a faster strategy that maintains acceptably
accurate results.

We consider shorter measurement times using the 2.25 multiplier experiments
from the previous section. Each experiment ran for 60 seconds, but we can
suppose it ran for a shorter time and take the median.
We emulate 10, 20, 30, and 60 second median
strategies in this way and present the accuracy of the results in
Figure~\ref{fig:stop-cond}.
The range of results generally gets larger as measurement times decreases.
Interestingly the 30 second
median strategy has the smallest range with all results falling between 0.84
and 1.01 times their ground-truth Tor capacity. For this reason and because it
is a reasonable balance between time-to-result and accuracy, we choose a
measurement length of 30 seconds.
See Appendix~\ref{sec:alt-msm-strat} for a description of more complex measure
strategies we considered but that performed worse.

\subsection{Measurement Strategies}\label{sec:alt-msm-strat}

In \S~\ref{sec:stop-cond} we chose a 30 second measurement duration, after which
\tool{} would take the median per-second throughput as the result of a
measurement. We now consider two other strategies.

\paragraph{Median with ignored lead time}
In this strategy \tool{} collects per-second throughput for a set duration $d$
as before, but \tool{} ignores the first $i$ seconds such that it takes the
median of the last $d - i$ seconds of data. The motivation is to avoid an
initial slow period due to TCP slow start, but because \tool{} uses many
sockets, it generally achieves its maximum throughput immediately (e.g.
Figure~\ref{fig:bg-traffic-10}). Consequently, a measurement of duration $d$ with
this strategy performs about as well as a simple median strategy of duration
$d - i$.

\paragraph{Dynamic Duration}
We also consider strategies with a dynamic measurement duration. As per-second
throughput results are gathered, the data points obtained thus far are viewed
in a series of windows. When the median throughput between each window changes
less than some factor, the measurement is stopped and the last window's median
throughput is the result. This strategy has similar motivations as the previous
one, but generally produces worse results than simple medians.

\subsection{Error bounds}\label{sec:error-bounds}
All parameters are now determined, but error bounds on what is considered an
accurate result are yet to be specified.
Recall from
\S~\ref{subsec:measuring-relay}
that \tool{} assigns some factor of measurer capacity
$f = m(1+\epsilon_2)/(1-\epsilon_1)$ times the existing capacity estimate for a
relay, where $m$ is the base multipler and $\epsilon_1$ and $\epsilon_2$ are
the error bounds.
Given the chosen measurement duration, 30
seconds, and its minimum/maximum fraction of ground-truth Tor capacity of
0.84/1.01 in Figure~\ref{fig:stop-cond}, we choose error bounds of
$\epsilon_1=\errlow{}$ and $\epsilon_2=\errhigh{}$ to leave a little room for
additional variation when we evaluate \tool's accuracy in
\S~\ref{sec:accuracy} and Appendix~\ref{sec:concurrent-measurements}.

\section{Concurrent Internet Measurements} \label{sec:concurrent-measurements}

A \tool{} deployment measures multiple relays at once in order to speed up the
rate at which it can measure the entire network, as described in
\S~\ref{subsec:measuring-network}. To evaluate \tool accuracy
when measuring relays concurrently, we
first establish ground-truth Tor capacity of relays limited with
\textsf{RelayBandwidthRate} to 100, 200, and 400 Mbit/s with the same method as
in \S~\ref{sec:environment}: we find ground truth
to be 94.2 Mbit/s, 191 Mbit/s, and 393 Mbit/s, respectively.  We then run
experiments with three sets of throughput-limited relays on US-SW: eight 100
Mbit/s relays, four 200 Mbit/s relays, and two 400 Mbit/s relays.  To perform
the measurements we choose the measurers US-E and NL as together they have the
smallest combined measurer capacity ($941+1611=2552$ Mbit/s, see
Table~\ref{tab:host-summary}) greater than the capacity necessary to
measure 800 Mbit/s of relay capacity at once
($800\cdot2.25[1+\errhigh]/[1-\errlow] = 2362.5$ Mbit/s, see
\S~\ref{subsec:measuring-relay}).  US-E and NL measure eight, four, or two
relays at once for 30 seconds; the result is the median per-second throughput.

\begin{table}
\centering
\footnotesize
\begin{threeparttable}
\captionsetup{skip=0pt} %
\caption{\tool estimates during concurrent measurement$^{\star}$}
\label{tab:concurrent}
\begin{tabular}{ccc|cc}
\toprule
\textbf{Limit} & \textbf{Capacity} & \textbf{Relays} & \textbf{Absolute} & \textbf{Relative}\\
(Mbit/s) & (Mbit/s) & (\#) & (Mbit/s) & (\%) \\
\midrule
100 & 94.2 & 8 & [87.6, 98.9] & [93, 105]\\
200 & 191  & 4 & [162, 185] & [85, 97]\\
400 & 393  & 2 & [307, 393] & [78, 100]\\
\bottomrule
\end{tabular}
\begin{tablenotes}
\footnotesize
\item[$\star$] Relays were run on US-SW, and measurers were run US-E and NL
\end{tablenotes}
\end{threeparttable}
\end{table}

The experiments and concurrent measurement results are summarized in
Table~\ref{tab:concurrent}. We observe that \tool{} can measure
accurately (within $\epsilon_1=\errlow{}$ and $\epsilon_2=\errhigh{}$)
in all but one case: one 400 Mbit/s relay measurement fell below
$\epsilon_1=\errlow{}$ by a relative factor of 0.02. The remaining
estimates are consistent with our accuracy results from measuring
single 250~Mbit/s relays at the $m = 2.25$ multiplier
(Figures~\ref{fig:verify} and~\ref{fig:mults}). Therefore, we conclude
that measuring relays concurrently does not negatively effect
\tool{}'s accuracy.

\end{document}